\documentclass[%
 reprint,
 amsmath,amssymb,
 aps,
]{revtex4-2}

\usepackage{lmodern}
\usepackage[T1]{fontenc}
\usepackage[latin9]{inputenc}
\usepackage{amsmath}
\usepackage{amsthm}
\usepackage{amssymb}
\usepackage{graphicx}
\usepackage{xcolor}
\usepackage{soul}

\usepackage{graphicx}
\usepackage{dcolumn}
\usepackage{bm}

\makeatletter
\usepackage[T1]{fontenc}
\usepackage{amsmath}
\usepackage{amsthm}
\usepackage{amssymb}
\usepackage{graphicx}
\usepackage{tikz-feynman}
\tikzfeynmanset{compat=1.1.0}

\usepackage{graphicx}
\usepackage{float}
\usepackage{xcolor}
\usepackage[hypertexnames=false]{hyperref}
\hypersetup{
	breaklinks = true,
    colorlinks = true,
    citecolor = {blue},
	urlcolor = {blue},
	linkcolor = {blue}
}

\makeatother

\usepackage{babel}
\begin{document}
\title{Self-Consistent Dynamics of Electron Radiation Reaction via Structure-Preserving Geometric Algorithms for Coupled Schr\"odinger-Maxwell Systems}

\author{Jacob Matthew Molina}
\email{jmmolina@princeton.edu }
\affiliation{Princeton Plasma Physics Laboratory, Princeton University, Princeton,
NJ 08540}
\affiliation{Department of Astrophysical Sciences, Princeton University, Princeton,
NJ 08540}

\author{Hong Qin}
\email{hongqin@princeton.edu}
\affiliation{Department of Astrophysical Sciences, Princeton University, Princeton,
NJ 08540}

\begin{abstract} Classically, a charged particle in a magnetic field emits radiation, losing momentum and experiencing the Abraham-Lorentz (AL) / Landau-Lifshitz (LL) radiation reaction (RR) force. However, at atomic scales and outside the range of their applicability, the AL/LL equations fail and RR destroys the coherent state of an electron---undermining the very concept of a RR force. This process can be described by the coupled Schr\"odinger-Maxwell (SM) system under appropriate limits, but the system's nonlinear complexity has long limited its use in purely analytical studies. We present geometric structure-preserving algorithms for the SM system that preserve gauge invariance, symplecticity, and unitarity on the discrete space-time lattice, which are implemented in our Structure-Preserving scHrodINger maXwell (SPHINX) code. By constructing coherent states from the Landau levels, SPHINX simulates the fully-coupled nonlinear dynamics of an electron coherent state, the energy partition evolution, and decoherence/relaxation of the electron wave packet in time due to RR. These simulations indicate that, in an external magnetic field, an electron prepared in an atomic-scale coherent state can radiate strongly, rapidly losing orbital coherence and dispersing into a decoherent wave packet. Additionally, we also present the fully-coupled nonlinear evolution of the non-degenerate ground- and first-excited Landau levels themselves to understand how the coupled SM system modifies the well-known ideal (i.e., Schr\"odinger-only) dynamics of the Landau Levels. With appropriate boundary conditions, simulation shows that the Landau levels are renormalized into stationary dressed eigenstates with constant electromagnetic and kinetic energies, providing a natural basis for the coupled electron-photon system. This opens a new computational window into RR physics and advances modeling of extreme-field phenomena in fusion plasmas, astrophysics, and next-generation laser experiments.
\end{abstract}
\maketitle

\section{Introduction}
A charged particle in a uniform magnetic field undergoes cyclotron motion, constantly accelerating and emitting radiation as it does \cite{Griffiths1998}. Conservation of momentum requires there exist an equal-and-opposite reaction force as the radiation is emitted: what is this force? While a seemingly simple question on its face, the nature of this force -- the so-called Radiation Reaction (RR) force -- has been a contentious question in physics since its inception. RR solutions have historically been garnered through a variety of different methods \cite{Lorentz1936CollectedPapers,Abraham1904,Dirac1938,Landau2013ClassicalTheoryFields}. The non-relativistic Abraham-Lorentz (AL) force can be heuristically derived from the Larmor formula, and can be extended to relativistic velocities through a similar argument instead beginning from the Lie\'nard radiation formula \cite{Dirac1938}. However, the AL force has well-known theoretical limitations; failing at atomic scales below the Compton wavelength \cite{Rohrlich1997}. Furthermore, the AL equation fails to provide an accurate description of the instantaneous particle radiation / energy loss in even as simple a case as cyclotron motion \cite{Griffiths1998}. Fully quantum and relativistic RR forces derived from quantum field theory are beset by unseemly fundamental difficulties such as runaway solutions that predict an exponentially increasing RR force in which the classical point charge retro-causally accelerates prior to the application of the force \cite{Griffiths1998}. In the limit of a weak radiation damping force, the Landau-Lifshitz (LL) radiation damping force avoids these  issues but is limited in its applicability \cite{Landau2013ClassicalTheoryFields}. Neither the AL nor the LL models are first principles calculations. 

Historically, radiation reaction has been invoked to invalidate the classical picture of an electron gyrating around a nucleus at atomic scales: a classical charged particle in a Kepler orbit would radiate away its kinetic energy in roughly 10 picoseconds and spiral into the nucleus \cite{OlsenMcDonald2005}. A quantum description is therefore essential; an electron can be prepared in a coherent state, the quantum analogue of a classical orbit, even at atomic scales. How does such a classical-like coherent quantum state evolve under radiation self-consistently? This is one of the main questions addressed in the present study.

The RR problem belongs to the broader class of self-force / self-field problems, whereby particles interact with their own self-generated fields \cite{Griffiths1998,Jackson1998}. The fundamental difficulty of RR forces stems from the fact that an electron is neither a classical point particle nor a classical extended rigid body, cases for which the RR problem has been solved within classical electrodynamics \cite{Rohrlich2000,Rohrlich2007,Spohn2023}. Instead, an electron is described by a quantum wave function in spacetime, governed by the Dirac equation, and the dynamics of this wave function are coupled to that of the electromagnetic field (photons) governed by Maxwell's equations. We note that radiation reaction for an accelerating electron is not well represented as one or a few discrete QED scattering events. It is a cumulative effect arising from a large number of electron-photon interactions. The appropriate framework is the pre-quantized Dirac-Maxwell system, which captures the tree-level dynamics of the underlying QED \cite{shi2018}. What the Dirac-Maxwell equations, viewed as a spacetime PDE system, do not include are loop-level QED corrections. These effects are not the primary concern for the radiation-reaction physics in plasmas considered in the present study.

One strategy to developing a better understanding of RR processes can be garnered by self-consistently evolving the fields of electrons and photons.  In the regime of interest, the electrons are nonrelativistic: their typical kinetic energies and potential-energy variations are small compared with their rest energy. In this low-energy limit the Dirac equation reduces to the Pauli equation, and, when spin-dependent effects are not essential to the phenomena we study, it further reduces to the  Schr\"odinger equation with relative corrections of order  $(v/c)^2$.  Thus, we adopt the Schr\"odinger-Maxwell (SM) system in the present study as a quantum system that self-consistently evolves the fields of particles and photons and provides an important perspective on the RR processes.  

The SM system is growing in importance as high-field physics is becoming increasingly more relevant to active areas of research such as high energy density physics, controlled nuclear fusion, and experimental/laboratory astrophysics \cite{guan2010phase, DiPiazza2012,Liu2014fate,Hirvijoki2015,shi2016,Breuer2000,Zhang2020,shi2021,Qu2021,Qu2022,AlNaseri2023,Griffith2024, Los2025,AlNaseri2020,AlNaseri2023, AlNaseri2025}.  Particularly, the advent of ultra high-intensity lasers has already began to presage the growing need to understand RR in such systems \cite{Ridgers2017,Cole2018,Poder2018,Gong2019, Mishra2022,Los2025, Redshaw2025}.  Purely analytical studies in this vein, however, are prohibited by the non-linear nature of the coupled SM system. 

We adopt geometric structure-preserving algorithms for the SM system that preserve gauge invariance, symplecticity, and unitarity on the discrete space-time lattice \cite{chen2017canonical}, and implement these algorithms into our MATLAB-based \textbf{S}tructure-\textbf{P}reserving Scr\textbf{h}\"od\textbf{in}ger-Ma\textbf{x}well (\textbf{SPHINX}) code. By constructing a coherent state from the Landau level eigenstates of the Hamiltonian for an electron in a uniform magnetic field, we can leverage SPHINX to simulate the nonlinear dynamics of a (spinless) electron coherent state self-consistently with the evolution of the electromagnetic fields.  Our simulations show that, in an external magnetic field, an electron initially prepared in an atomic-scale coherent state can radiate a substantial amount of electromagnetic energy, and the classical-like orbit rapidly loses coherence and disperses into a decoherent state. In contrast, when radiation reaction is treated self-consistently with appropriate boundary conditions, the standard Landau-level eigenstates are modified. These modified eigenstates maintain constant levels of electromagnetic and kinetic energy and provide a natural basis for describing electron state together with its self-consistent electromagnetic field. 

This work is structured as follows. We begin by first deriving the algorithms implemented into SPHINX in section \ref{dynamicsSec}, beginning with a review of the dynamics of the continuous SM system in subsection \ref{contDynamSec}, then turning to the dynamics admitted by the discrete SM system in subsection \ref{discDynamSec}, and then finally furnishing our algorithms in subsection \ref{algoSection}. In section \ref{simSec} we present our simulations studying the uncoupled (subsection \ref{sec:staticSims}) and coupled (subsection \ref{sec:coupledSims}) dynamics of the electron coherent state. In the same section, we also present the ground and first-excited states of the Landau level eigenmodes (subsection \ref{sec:llSims}); the theory underlining our derivation of the Landau levels and the coherent state we use can be found in the appendix section \ref{llAndCsTheoryAppendix}. Finally, we provide a final discussion of and conclusions drawn from our results in section \ref{concSec}.

\section{Dynamics of the Schr\"odinger-Maxwell System} \label{dynamicsSec}
In this section we will derive the geometric structure-preserving algorithms used to numerically solve the Schr\"odinger-Maxwell (SM) system of equations, given by:
\begin{align} 
    i \hbar \frac{\partial}{\partial t} \psi =& \mathcal{H} \psi \label{SE}, \\ 
        \partial_{\mu} F^{\mu \nu} =& \mu_{0} J^{\mu} \label{ME}.
\end{align}
The Schr\"odinger equation (SE) is given by Eq.\,\eqref{SE}, where the Hamiltonian operator ($\mathcal{H}$) given by $\mathcal{H} \equiv {\left({\textbf{P}} - q \textbf{A} \right)^2}/{2m} $ for which the momentum operator (${\textbf{P}}$) is given by ${\textbf{P}} \equiv -i \hbar \mathbf{\nabla}$. In what follows, we adopt the temporal gauge (i.e., $\phi = 0$).  Maxwell's equations (ME) are given in the geometric form in Eq.\,\eqref{ME}, where $F \equiv  \partial^{\mu} \textbf{A}^{\mu} - \partial^{\nu} \textbf{A}^{\mu} $ is the electromagnetic tensor and $\textbf{A}$ is the vector potential satisfying $\nabla \times \mathbf{A} = \mathbf{B}$ and $\dot{\mathbf{A}} = - \mathbf{E}$, where $\mathbf{B}$ and $\mathbf{E}$ are the magnetic and electric fields respectively. Here, $J^{\mu}$ is the conserved 4-current given by $J^{\mu} \equiv i \left[ \psi^{*} D^{\mu} \psi - \psi \left( D^{\mu} \psi \right)^{*} \right]$ with $D^{\mu} \equiv \partial_{\mu} + i\frac{q}{\hbar}\textbf{A}_{\mu}$ is the gauge co-variant derivative assuming a metric signature $\left( +,-,-,- \right)$. In this form $J^{0} = cq |\psi|^{2} \equiv c \rho$ and $J^{k} \equiv \mathcal{J}^{k} = \frac{q}{2m} \left[ \psi^{*} (i\hbar D^{k} \psi) - \psi \left( i \hbar D^{k} \psi \right)^{*} \right]$ for $k \in \{1,2,3\}$, where $\rho$ is the charge density and $\mathcal{J}$ is the electromagnetic current for  that couples the SE to ME. 

In what follows, we will begin by first reviewing the derivation for the dynamics of the continuous SM system before deriving the dynamic equations for the discrete SM system. Finally, from the dynamic equations for the discrete system we utilize a symplectic splitting scheme that will furnish our geometric structure-preserving algorithms for evolving the discrete SM system.  The theoretical and algorithmic formulation presented here is same as that in Ref.\,\cite{chen2017canonical}, but streamlined for a single matter field without non-self-consistent external potential for the study of radiation reaction effect. More broadly, the algorithms adopted belong to the class of structure-preserving geometric algorithms in plasma physics \cite{squire2012geometric,Xiao2013,xiao2015explicit,he2015Hamiltonian,he2016hamiltonian,qin2016canonical,kraus2017gempic,burby2017finite,Morrison2017,xiao2017local,Xiao2018review,Xiao2019field,Xiao2021Explicit,Glasser2020,Wang2021,Kormann2021,Perse2021,Glasser2022,CamposPinto2022,Burby2023,Qin2025a}  that have been developed in the last decade. 

\subsection{Dynamics of the Continuous Schr\"odinger-Maxwell System} \label{contDynamSec}
Let us begin by reviewing the dynamics admitted by the continuous SM system. The Hamiltonian describing the SM system is composed of the individual Hamiltonians for the quantum and electromagnetic sub-systems, denoted here by $H_{qm}$ and $H_{em}$ respectively 

\begin{align} 
        H = & H_{qm} + H_{em} \label{H} ,\\
        H_{qm} = & \int d^3 x \left[ \psi^{*} \mathcal{H} \psi \right] \label{Hqm}, \\
        H_{em} = & \int d^3 x \left[ \frac{|\nabla \times \textbf{A}|^2}{2\mu_{0}} + \frac{\epsilon_{0} \dot{\mathbf{A}}^2}{2} \right]. \label{Hem}
\end{align}
The symplectic 2-form of the Schr\"odinger field, $\Omega_{qm}$, and concomitant canonical Poisson bracket structure for the quantum sub-system \cite{chen2017canonical}, $\left\{ F,G \right\}_{qm}$ , are given by 
\begin{align} 
    \Omega_{qm} = & \hbar \int d^3 x [d\psi_{R} \wedge d\psi_I] ,\\
    \left\{ F,G \right\}_{qm} = & \frac{1}{\hbar}\int d^{3}x\left[ \frac{\delta F}{\delta \psi_{R}}\frac{\delta G}{\delta \psi_{I}} -\frac{\delta G}{\delta \psi_{R}}\frac{\delta F}{\delta \psi_{I}} \right],
\end{align}
where F and G are understood to be functionals of our pairs of canonical variables $(\psi_{R},\psi_{I})$, and $\delta F / \delta \mathcal{X}$ denote variational derivatives with respect to a variable $ \mathcal{X}$. By decomposing $\psi = \frac{1}{\sqrt{2}} \left( \psi_{R} +  i \psi_{I} \right)$ and $\mathcal{H} = \mathcal{H}_{R} + i\mathcal{H}_{I}$ we can rewrite Eqs.\,\eqref{SE} and \eqref{Hqm} in terms of $\psi_{R}$, $\psi_{I}$, $\mathcal{H}_{R}$, and $\mathcal{H}_{I}$ as

\begin{align} \label{SM_Ham}
    \hbar \frac{\partial}{\partial t} \begin{pmatrix}
        \psi_{R} \\ \psi_{I}
    \end{pmatrix} =& \mathcal{H}_{RI}\begin{pmatrix}
        \psi_{R} \\ \psi_{I}
    \end{pmatrix},\\
     \mathcal{H}_{RI}\equiv&\begin{pmatrix}
      \mathcal{H}_{I} & \mathcal{H}_{R} \\ -\mathcal{H}_{R} & \mathcal{H}_{I}
    \end{pmatrix} ,
\end{align}
and
\begin{align}
    H_{qm} = &  \int d^3x  \left( -\frac{\hbar^2}{2m} \text{Re}\left[ \psi^* \nabla^2 \psi \right] \right. \label{Hqm_ref} \\
    &\left. - \frac{q \hbar}{m} \text{Im}\left[ \psi^* \textbf{A} \cdot \nabla \psi \right] + \frac{q^2}{2m} |\textbf{A}|^2 |\psi|^2\right)  \notag \\
    = ~ & \frac{1}{2}\int d^3x \left[ \psi_{R} \mathcal{H}_{R} \psi_{R} + \psi_{I} \mathcal{H}_{I} \psi_{I} \right. \\
    & \left. + \psi_{I} \mathcal{H}_{I} \psi_{R} - \psi_{R} \mathcal{H}_{R} \psi_{I}\right] \notag
\end{align}

Unitarity of $\psi$ is equivalent to the real--imaginary splitting $(\psi_R,\psi_I)$ being both symplectic and orthogonal \cite{deGosson06, Qin2013PRL-GKVS,Qin2013PRL2,Gosson2016,Qin2019LH}, which, in finite dimension, means
\begin{equation}
SO(2n)\cap Sp(2n,\mathbb{R})\cong U(n).
\end{equation}
Equivalently, the matrix operator $\mathcal{H}_{RI}$ on the right-hand side of Eq.~\eqref{SM_Ham} lies in the Lie algebra of the symplectic-rotation group.  The structure-preserving algorithm presented will preserve exactly this structure on the discrete level. 

Assuming $(\psi_{R},\psi_{I})$ are either fixed or periodic at the boundaries, we can compute the variation of $\delta H_{qm}$ via integration by parts as:
\begin{align} \label{delta_Hqm}
    \delta H_{qm} =  \frac{1}{2m} \int d^{3}x  & \left[ \left(-\hbar^2 \nabla^2 + q^2\textbf{A}^2 + 2V \right) \psi_{R} \right. \\
    & \left. - \left(2q\hbar\textbf{A} \cdot \nabla + \nabla \cdot q\hbar\textbf{A} \right) \psi_{I} \right] \delta \psi_{R} \notag \\ 
     + & \left[ \left(-\hbar^2 \nabla^2 + q^2\textbf{A}^2 + 2V \right) \psi_{I} \right. \notag \\
    & + \left. \left(2q\hbar\textbf{A} \cdot \nabla + \nabla \cdot q\hbar\textbf{A} \right) \psi_{R} \right] \delta \psi_{I} \notag \\ 
    + & \left[ q\hbar \left( \psi_{I} \nabla \psi_{R} -  \psi_{R} \nabla \psi_{I} \right) \right. \notag \\ 
    & + \left. q^2  \left(\psi_{R}^{2}+ \psi_{I}^{2}\right)\textbf{A} \right] \cdot \delta \textbf{A}. \notag
\end{align}

In the case of the electromagnetic sub-system, the symplectic 2-form of the electromagnetic field, $\Omega_{em}$, and the concomitant canonical Poisson bracket structure \cite{Marsden1982Hamiltonian,qin2016canonical}, $\left\{ F,G \right\}_{em}$,  are given by
\begin{align}
    \Omega_{em}= & \int d^3 x [d\textbf{A} \wedge d{\mathbf{Y}}],\\
    \left\{ F,G \right\}_{em} = & \int d^3 x \left[\frac{\delta F}{\delta \textbf{A}}\frac{\delta G}{\delta \textbf{Y}} - \frac{\delta G}{\delta \textbf{A}}\frac{\delta F}{\delta \textbf{Y}}\right], 
\end{align}
where $\textbf{Y} \equiv \epsilon_{0} \dot{\mathbf{A}}$. In terms of $\textbf{A}$ and $\textbf{Y}$, Eq. \,\eqref{Hem} is given by
\begin{equation} 
        H_{em} = \frac{1}{2}\int d^3 x \left[ \frac{|\nabla \times \textbf{A}|^2}{\mu_{0}} + \frac{\textbf{Y}^2}{\epsilon_{0}} \right] \label{Hem_ref}.
\end{equation}
Under the same assumption that our fields are either fixed or periodic at the boundaries, it straight forward to prove that the variation in $H_{em}$ is given by
\begin{equation} \label{delta_Hem}
\delta H_{em}  = \int d^3x \left[ \frac{1}{\mu_{0}} \nabla \times \nabla \times \textbf{A} \cdot \delta \textbf{A} + \frac{\textbf{Y}}{\mu_{0}} \cdot \delta \textbf{Y}  \right].
\end{equation}
For our choice of canonical variables $(\psi_{R},\psi_{I},\textbf{A},\textbf{Y})$,  the full symplectic structure and full canonical Poisson bracket are then given by
\begin{align} 
    \Omega&=\Omega_{qm}+\Omega_{em},\\
    \left\{ F,G \right\} &= \left\{ F,G \right\}_{qm} + \left\{ F,G \right\}_{em} \label{contPois},
\end{align}
 and Eqs.\,\eqref{delta_Hqm}, \eqref{delta_Hem} and \eqref{contPois} then produce the following system of coupled non-linear partial differential equations describing the time-evolution of the SM system:

\begin{align}
    \Dot{\psi_{R}}  =  \left\{ \psi_{R},H \right\} = ~& \frac{q}{2m} \nabla \cdot \textbf{A}\psi_{R} + \frac{q}{m} \textbf{A} \cdot \nabla \psi_{R} \label{psiR_cont}  \\
    & + \frac{1}{2m \hbar} \left( -\hbar^2 \nabla^{2} + q^{2} \textbf{A}^{2} \right) \psi_{I}  \notag, \\
    \Dot{\psi_{I}}  = \left\{ \psi_{I},H \right\} =  ~& \frac{q}{2m} \nabla \cdot \textbf{A}\psi_{I} + \frac{q}{m} \textbf{A} \cdot \nabla \psi_{I} \\
    & + \frac{1}{2m \hbar} \left(\hbar^2 \nabla^{2} - q^{2} \textbf{A}^{2} \right) \psi_{R} \notag, \\
    \Dot{\textbf{A}} = \left\{ \textbf{A},H \right\} = ~& \frac{1}{\epsilon_{0}} \textbf{Y}, \\
    \Dot{\textbf{Y}}  = \left\{ \textbf{Y} ,H \right\} = ~& \mathcal{J} - \frac{1}{\mu_{0}} \nabla \times \nabla \times \textbf{A} \label{Y_cont} \\
    \mathcal{J} = ~& \frac{q}{2m} \left[ \hbar \left( \psi_{R} \nabla \psi_{I} - \psi_{I} \nabla \psi_{R} \right) \right. \\
    & ~~~~~~ \left. - \left( \psi_{R}^2 + \psi_{I}^{2} \notag \right)q\textbf{A} \right].
\end{align}

\subsection{Dynamics of the Discrete Schr\"odinger-Maxwell System} \label{discDynamSec}
We now set out to derive the dynamics of the discrete system. We choose to deposit our fields onto an Eulerian grid of $M$ grid points as \cite{qin2016canonical,chen2017canonical}

\begin{equation}
    \begin{split} \label{deposition}
            \psi_{R}\left(t,\textbf{x} \right) = & \sum_{J = 1}^{M} \psi_{RJ} \left( t \right) \theta(\textbf{x} - \textbf{x}_{J}) ,\\
            \psi_{I}\left(t,\textbf{x} \right) = & \sum_{J = 1}^{M} \psi_{IJ} \left( t \right) \theta(\textbf{x} - \textbf{x}_{J}) , \\
            \textbf{A}\left(t,\textbf{x} \right) = &\sum_{J = 1}^{M} \textbf{A}_{J} \left( t \right) \theta(\textbf{x} - \textbf{x}_{J})  ,\\
            \textbf{Y}\left(t,\textbf{x} \right) = & \sum_{J = 1}^{M} \textbf{Y}_{J} \left( t \right) \theta(\textbf{x} - \textbf{x}_{J}), 
    \end{split}
\end{equation}
where $J$ references the 3D grid point $J = (i,j,k)$, and the distribution function $\theta \left(\textbf{x} - \textbf{x}_{J} \right) $ is defined as  
\begin{equation} \label{theta}
\theta \left(\textbf{x} - \textbf{x}_{J} \right) = \begin{cases} 1, & |x - x_{J}| < \frac{\Delta x}{2},|y - y_{J}| < \frac{\Delta y}{2}, \\ & |z - z_{J}| < \frac{\Delta z}{2} \\ 0, &\text{elsewhere}. \end{cases}. 
\end{equation}

By applying $\int d^3 x \theta(\textbf{x}-\textbf{x}_{K})[...]$ to, for example, the grid deposition scheme for $\textbf{A}(t,\textbf{x})$ given in Eq.\,\eqref{deposition} and leveraging the orthogonality of $\theta(\textbf{x}-\textbf{x}_{K})$ and $\theta(\textbf{x}-\textbf{x}_{J})$, one can prove that $\frac{\delta \textbf{A}_{J}}{\delta \textbf{A}} \equiv \frac{1}{\Delta V} \theta \left( \textbf{x} - \textbf{x}_{J} \right)$, where $\Delta V \equiv \Delta x \Delta y \Delta z$ is the volume of each grid cell. With this in hand, we define the discrete variational derivative
\begin{equation} \label{discVarDeriv}
    \frac{\delta F}{\delta \textbf{A}} = \sum_{J = 1}^{M} \frac{\delta \textbf{A}_{J}}{\delta \textbf{A}} \frac{\partial F}{\partial \textbf{A}_{J}} = \sum_{J = 1}^{M} \frac{1}{\Delta V} \theta \left( \textbf{x} - \textbf{x}_{J} \right) \frac{\partial F}{\partial \textbf{A}_{J}},
\end{equation}
and discretize the canonical Poisson bracket structure as
\begin{align} \label{discPoisBrac}
    \left\{F,G \right\}_{d} =  \sum_{J = 1}^{M} &\frac{1}{\Delta V} \left[  \frac{1}{\hbar} \left( \frac{\partial F}{\partial \psi_{RJ}}\frac{\partial G}{\partial \psi_{IJ}} - \frac{\partial G}{\partial \psi_{RJ}}\frac{\partial F}{\partial \psi_{IJ}}\right) \right. \\ 
     & + \left. \frac{\partial F}{\partial \textbf{A}_{J}}\frac{\partial G}{\partial \textbf{Y}_{J}} - \frac{\partial G}{\partial \textbf{A}_{J}}\frac{\partial F}{\partial \textbf{Y}_{J}} \right] \notag .
\end{align}

In what follows, the subscript $d$ will be added to denote that a quantity or operator is discrete and distinct from the continuous version outlined in the previous sub-section. The Hamiltonian functional for the SM system is discretized as

\begin{align}
        H_{d} = H_{dqm} + & H_{dem},   \\
        H_{dqm} = \frac{1}{2m} \sum_{J}^{M} & \left[ -\frac{1}{2} \hbar^{2} \psi_{RJ} \left( \nabla^{2}_{d} \psi_{R} \right)_{J} - \frac{1}{2} \hbar^{2} \psi_{IJ} \left( \nabla^{2}_{d} \psi_{I} \right)_{J} \right. \notag \\
         - q\hbar \psi_{RJ} & \textbf{A}_{J} \cdot \left( \mathbf{\nabla}_{d} \psi_{I}  \right)_{J}   +  q\hbar \psi_{IJ} \textbf{A}_{J} \cdot \left( \mathbf{\nabla}_{d} \psi_{R} \right)_{J}  \notag \\
        + & \left. \frac{1}{2}  q^{2} A^{2}_{J} \left( \psi^{2}_{RJ} + \psi^{2}_{IJ} \right) \right]  \Delta V , \\
         H_{dem} = \frac{1}{2} & \sum_{J}^{M} \left[ \frac{1}{\epsilon_{0}} \textbf{Y}^{2}_{J} + \frac{1}{\mu_{0}} \left( \mathbf{\nabla}_{d} \times \textbf{A}\right)^{2}_{J} \right] \Delta V. 
\end{align}

In this work we utilize a backwards differencing method for our derivatives. Operating on some scalar field $\psi_{J} \equiv \psi_{i,j,k}$ and some vector field $\textbf{A}_{J} \equiv \textbf{A}_{i,j,k} = \langle A_{x,J},A_{y,J},A_{z,J} \rangle $, our discrete differential operators take the following form \cite{qin2016canonical,chen2017canonical}: 
\begin{align} \label{divScheme}
\left( \nabla_{d} \cdot \textbf{A} \right)_{J} & \equiv \frac{{A_{x}}_{i,j,k} - {A_{x}}_{i-1,j,k}}{\Delta x} \notag \\
& ~~~ + \frac{{A_{y}}_{i,j,k} - {A_{y}}_{i,j-1,k}}{\Delta y} \ \\ 
& ~~~ + \frac{{A_{z}}_{i,j,k} - {A_{z}}_{i,j,k-1}}{\Delta z} \notag ,\\
\label{curlScheme}
\left( \nabla_d \times \textbf{A} \right)_{J} & \equiv \begin{pmatrix}
            \frac{{A_{z}}_{i,j,k} - {A_{z}}_{i,j-1,k}}{\Delta y} - \frac{{A_{y}}_{i,j,k} - {A_{y}}_{i,j,k-1}}{\Delta z} \\
            \frac{{A_{x}}_{i,j,k} - {A_{x}}_{i,j,k-1}}{\Delta z} - \frac{{A_{z}}_{i,j,k} - {A_{z}}_{i-1,j,k}}{\Delta x} \\
            \frac{{A_{y}}_{i,j,k} - {A_{y}}_{i-1,j,k}}{\Delta x} - \frac{{A_{x}}_{i,j,k} - {A_{x}}_{i,j-1,k}}{\Delta y} \\
        \end{pmatrix},\\
\label{gradScheme}
\left( \nabla_{d} \psi \right)_{J} & \equiv \begin{pmatrix}
            \frac{\psi_{i,j,k} - \psi_{i-1,j,k}}{\Delta x} \\
            \frac{\psi_{i,j,k} - \psi_{i,j-1,k}}{\Delta y} \\
            \frac{\psi_{i,j,k} - \psi_{i,j,k-1}}{\Delta z} \\
        \end{pmatrix} ,\\
\label{laplaceScheme}
        \left( \nabla_d^{2} \psi \right)_{J} & \equiv \frac{\psi_{i,j,k} - 2 \psi_{i-1,j,k} + \psi_{i-2,j,k}}{\Delta x^{2}} \\
        & ~~~ + \frac{\psi_{i,j,k} - 2 \psi_{i,j-1,k} + \psi_{i,j-2,k}}{\Delta y^{2}} \notag \\
        & ~~~ + \frac{\psi_{i,j,k} - 2 \psi_{i,j,k-1} + \psi_{i,j,k-2}}{\Delta z^{2}}. \notag
\end{align}

It follows that, as with the continuous system, we can derive dynamic equations for the discrete SM system through our discrete Poisson bracket structure:
\begin{align} \label{discDyn}
        \Dot{\psi_{RJ}} & = \left\{ \psi_{RJ},H_{d} \right\}_{d} \\
        & = \frac{q}{2m}  \textbf{A}_{J} \cdot \left( \nabla_{d} \psi_{R}\right)_{J} - \frac{q}{2m} \sum_{K = 1}^{M} \psi_{RK} \textbf{A}_{K} \cdot \frac{\partial \left( \nabla_{d} \psi_{I}\right)_{K}}{\partial \psi_{IJ}}  \notag \\
        & ~~~ - \frac{\hbar}{4m} \left( \nabla_{d}^{2} \psi_{I} \right)_{J} - \frac{\hbar}{4m} \sum_{K=1}^{M} \psi_{IK} \frac{\partial \left( \nabla^{2}_{d} \psi_{I} \right)_{K}}{\partial \psi_{IJ}}  \notag \\
        & ~~~ + \frac{1}{\hbar} \frac{q^2}{2m}\textbf{A}_{J}^{2} \psi_{IJ} \notag, \\
        \Dot{\psi_{IJ}} & = \left\{ \psi_{IJ},H_{d} \right\}_{d} \\
        & = \frac{q}{2m} \textbf{A}_{J} \cdot \left( \nabla_{d} \psi_{I}\right)_{J} - \frac{q}{2m} \sum_{K = 1}^{M} \psi_{IK} \textbf{A}_{K} \cdot \frac{\partial \left( \nabla_{d} \psi_{R}\right)_{K}}{\partial \psi_{RJ}}  \notag ,\\
        & ~~~ + \frac{\hbar}{4m} \left( \nabla_{d}^{2} \psi_{R} \right)_{J} + \frac{\hbar}{4m} \sum_{K=1}^{M} \psi_{RK} \frac{\partial \left( \nabla^{2}_{d} \psi_{R} \right)_{K}}{\partial \psi_{RJ}}  \notag \\
        & ~~~ - \frac{1}{\hbar}\frac{q^2}{2m}\textbf{A}_{J}^{2}  \psi_{RJ} \notag ,\\
        \Dot{\textbf{A}_{J}} & = \left\{ \textbf{A}_{J},H_{d} \right\}_{d}  = \frac{1}{\epsilon_{0}} \textbf{Y}_{J},\\
        \Dot{\textbf{Y}}_{J} & = \left\{ \textbf{Y}_{J} ,H_{d} \right\}_{d} \label{discY} \\
        & = \mathcal{J}_{J} - \frac{1}{2\mu_{0}} \sum_{K=1}^{M} \frac{\partial}{\partial \textbf{A}_J}\left( \nabla_{d} \times \textbf{A} \right)_{K}^{2} \notag ,\\
        \mathcal{J}_{J} & = \frac{q}{2m} \left[ \hbar \left( \psi_{RJ} \left( \nabla_{d} \psi_{I} \right)_{J} - \psi_{IJ} \left( \nabla_{d} \psi_{R} \right)_{J} \right) \right. \\
        & ~~~~~~~~~~ \left. - \left( \psi_{RJ}^2 + \psi_{IJ}^{2} \right)\textbf{A}_{J} \right]. \notag 
\end{align}
Here, we can cast the second term in Eq.\,\eqref{discY} into a more familiar form by utilizing the fact that $\frac{1}{2}  \sum_{K=1}^{M} \frac{\partial}{\partial \textbf{A}_J}\left( \nabla_{d} \times \textbf{A} \right)_{K}^{2}  \equiv \left( \nabla_{d}^{T} \times \nabla_{d} \times \textbf{A} \right)_{J}$ for a properly-chosen discrete curl operator, $\mathbf{\nabla}_{d} \times \left[ ~~~~\right]$.

\subsection{Structure-Preserving Geometric Algorithm} \label{algoSection}
In this work, we employ a symplectic splitting scheme whereby we split $H_{d}$ into the separate quantum ($H_{dqm}$) and electromagnetic ($H_{dem}$) parts. In doing this, we derive dynamic equations under $H_{dqm}$ and $H_{dem}$ separately, and then combine the two solution maps to derive algorithms for the full discrete SM system. We choose to discretize our time derivatives via a symplectic mid-point scheme. Beginning with the simpler case of $\eta_{em}$ -- the map derived from $H_{dem}$ -- our dynamic equations can be derived as:
\begin{align}
        \Dot{\psi}_{RJ} & = \left\{ \psi_{RJ},H_{dem} \right\}_{d} = 0 \label{R_Hem},\\
        \Dot{\psi}_{IJ} & = \left\{ \psi_{IJ},H_{dem} \right\}_{d} = 0 \label{I_Hem},\\
        \Dot{\textbf{A}}_{J} & = \left\{ \textbf{A}_{J},H_{dem} \right\}_{d} = \frac{1}{\epsilon_{0}} \textbf{Y}_{J} \label{A_Hem} ,\\
        \Dot{\textbf{Y}}_{J} & = \left\{ \textbf{Y}_{J} ,H_{dem} \right\}_{d} = - \frac{1}{\mu_{0}}  \left( \nabla_{d}^{T} \times \nabla_{d} \times \textbf{A} \right)_{J} \label{Y_Hem}.
\end{align}
Naturally we find that $\psi$ does not evolve under $H_{dem}$, meaning  $\psi^{n+1} = \psi^{n}$ under $\eta_{em}$, where $n$ is used here to denote the time-step index and we  have additionally shifted this index $n \mapsto n+1$. In the case of the electromagnetic fields, we can rewrite Eqs.\,\eqref{A_Hem} and \eqref{Y_Hem} in the form

\begin{equation} \label{F_Hem}
        \frac{d}{dt} \begin{pmatrix}
        \textbf{A} \\ \textbf{Y}
    \end{pmatrix} = \Lambda \begin{pmatrix}
        \textbf{A} \\ \textbf{Y}
    \end{pmatrix},
\end{equation}
for a constant matrix operator $\Lambda$, which is an element in the symplectic algebra $\mathfrak{sp}(2Dim(M))$, where $Dim(M)$ is the dimension of grid points. Discretizing the time derivative and adopting a symplectic mid-point method  allows us to write Eq.\,\eqref{F_Hem} in the form

\begin{equation}
    \begin{split}
        \begin{pmatrix}
            \textbf{A} \\ \textbf{Y}
        \end{pmatrix}^{n} -     \begin{pmatrix}
            \textbf{A} \\ \textbf{Y}
        \end{pmatrix}^{n-1} = & \Delta t \Lambda \begin{pmatrix}
            \textbf{A} \\ \textbf{Y}
        \end{pmatrix}^{n-1/2} \\
         = & \frac{\Delta t}{2} \Lambda \left[ \begin{pmatrix}
            \textbf{A} \\ \textbf{Y}
        \end{pmatrix}^{n} + \begin{pmatrix}
            \textbf{A} \\ \textbf{Y}
        \end{pmatrix}^{n-1} \right],
    \end{split}
\end{equation}
where $\Delta t$ is the time-step. The solution for $(\textbf{A},\textbf{Y})^{n+1}$ expressed in term of the Cayley transform

\begin{equation}
     \begin{pmatrix}
        \textbf{A} \\ \textbf{Y}
    \end{pmatrix}^{n+1} = \text{Cay} \left[ \frac{\Delta t}{2}\Lambda\right] \begin{pmatrix}
        \textbf{A} \\ \textbf{Y}
    \end{pmatrix}^{n}.
\end{equation}

Here, $\text{Cay}\left[X\right] \equiv \left(1-X\right)^{-1} \left(1+X\right)$ is the Cayley transform of a matrix $X$.  It is a known property that if $X$ is in a Lie algebra,  its Cayley transform is in the corresponding Lie group \cite{Feng2010,Qin2013Boris}. Because $X$ is in the symplectic algebra,   $\text{Cay}\left[X\right]$ is a symplectic matrix. Thus, the map $(\textbf{A},\textbf{Y})^{n} \mapsto (\textbf{A},\textbf{Y})^{n+1}$ is necessarily symplectic. The $\psi^{n} = \psi^{n+1}$ map is trivially symplectic and unitary.  The symplectic and unitary  and map $\eta_{em}: \left(\psi,\textbf{A},\textbf{Y}\right)^{n} \mapsto \left(\psi,\textbf{A},\textbf{Y}\right)^{n+1}$ is then given by
\begin{equation}
\eta_{em}: \begin{cases}
         \psi^{n+1} = \psi^{n},\\
         \begin{pmatrix}
        \textbf{A} \\ \textbf{Y}
    \end{pmatrix}^{n+1} = \text{Cay} \left[ \frac{\Delta t}{2}\Lambda\right] \begin{pmatrix}
        \textbf{A} \\ \textbf{Y}
    \end{pmatrix}^{n}. \\
\end{cases}
\end{equation}

Similarly, to derive $\eta_{qm}$ we utilize $H_{dqm}$ to produce the following dynamic equations

\begin{align} 
        \Dot{\psi_{RJ}} & = \left\{ \psi_{RJ},H_{dqm} \right\}_{d} \label{R_Hqm} \\
        & = \frac{q}{2m} \textbf{A}_{J} \cdot \left( \nabla_{d} \psi_{R}\right)_{J} - \frac{q}{2m} \sum_{K = 1}^{M} \psi_{RK} \textbf{A}_{K} \cdot \frac{\partial \left( \nabla_{d} \psi_{I}\right)_{K}}{\partial \psi_{IJ}}  \notag \\
        & ~~~ - \frac{\hbar}{4m} \left( \nabla_{d}^{2} \psi_{I} \right)_{J} - \frac{\hbar}{4m} \sum_{K=1}^{M} \psi_{IK} \frac{\partial \left( \nabla^{2}_{d} \psi_{I} \right)_{K}}{\partial \psi_{IJ}}  \notag \\
        & ~~~ + \frac{1}{\hbar} \frac{q^2}{2m}\textbf{A}_{J}^{2}\psi_{IJ} \notag ,\\
        \Dot{\psi_{IJ}} & = \left\{ \psi_{IJ},H_{dqm} \right\}_{d} \label{I_Hqm} \\
        & ~~~ = \frac{q}{2m} \textbf{A}_{J} \cdot \left( \nabla_{d} \psi_{I}\right)_{J} - \frac{q}{2m} \sum_{K = 1}^{M} \psi_{IK} \textbf{A}_{K} \cdot \frac{\partial \left( \nabla_{d} \psi_{R}\right)_{K}}{\partial \psi_{RJ}}  \notag \\
        & ~~~ + \frac{\hbar}{4m} \left( \nabla_{d}^{2} \psi_{R} \right)_{J} + \frac{\hbar}{4m} \sum_{K=1}^{M} \psi_{RK} \frac{\partial \left( \nabla^{2}_{d} \psi_{R} \right)_{K}}{\partial \psi_{RJ}}  \notag \\
        & ~~~ - \frac{1}{\hbar} \frac{q^2}{2m}\textbf{A}_{J}^{2} \psi_{RJ} \notag, \\
        \Dot{\textbf{A}_{J}} & = 0 \label{A_Hqm},\\
        \Dot{\textbf{Y}}_{J} & = \left\{ \textbf{Y}_{J} ,H_{dqm} \right\}_{d} = \mathcal{J}_{J} \label{Y_Hqm}.
\end{align}
Equations \eqref{A_Hqm} and \eqref{Y_Hqm} lead to
\begin{align}
        \textbf{A}^{n+1} & = \textbf{A}^{n} , \\
        \textbf{Y}^{n+1} & = \textbf{Y}^{n} + \Delta t \mathcal{J} \left(\textbf{A}^{n},\psi_{R}^{n+1/2},\psi_{I}^{n+1/2} \right) \label{Y_Hqm2}. 
\end{align}
The RHS of Eq.\,\eqref{Y_Hqm2} can be computed by recognizing that $\mathcal{J}\left(\textbf{A},\psi_{R},\psi_{I}\right)|^{n = n+1/2} = \mathcal{J} \left(\textbf{A}^{n},\frac{\psi_{R}^{n+1}+\psi_{R}^{n}}{2},\frac{\psi_{I}^{n+1}+\psi_{I}^{n}}{2} \right)$ given that $\textbf{A}$ does not evolve under $\eta_{qm}$. 

Equations \eqref{R_Hqm} and \eqref{I_Hqm} can be expressed as
\begin{equation} \label{S_Hqm}
    \frac{d}{dt} \begin{pmatrix}
        \psi_{R} \\ \psi_{I}
    \end{pmatrix} = \Gamma \left( \textbf{A}^{n} \right) \begin{pmatrix}
        \psi_{R} \\ \psi_{I}
    \end{pmatrix},
\end{equation}
where $\Gamma \left( \textbf{A}^{n} \right)$  is a skew-symmetric and infinitesimal  symplectic matrix, i.e., $\Gamma \left( \textbf{A}^{n} \right)\in \mathfrak{so}(2Dim(M))\cap \mathfrak{sp}(2Dim(M),\mathbb{R})$. Similar to Eq.\,\eqref{F_Hem},  the corresponding one-step map can be expressed in terms of $\text{Cay}\left[\Gamma \left( \textbf{A}^{n} \right)\right]$ as
\begin{equation} \label{S_Hqm}
    \begin{pmatrix}
        \psi_{R} \\ \psi_{I}
    \end{pmatrix}^{n+1} = \text{Cay}\left[\Gamma \left( \textbf{A}^{n} \right)\right] \begin{pmatrix}
        \psi_{R} \\ \psi_{I}
    \end{pmatrix}^{n}.
\end{equation}
Invoking the property of Caylay transform again \cite{Feng2010,Qin2013Boris}, we note that $\text{Cay}\left[\Gamma \left( \textbf{A}^{n} \right)\right]$ is in the symplectic-rotation group, i.e., $\text{Cay}\left[\Gamma \left( \textbf{A}^{n} \right)\right]\in 
SO(2Dim(M))\cap Sp(2Dim(M),\mathbb{R})\cong U(Dim(M))$ \cite{deGosson06, Qin2013PRL-GKVS,Qin2013PRL2,Gosson2016,Qin2019LH},  because  $\Gamma \left( \textbf{A}^{n} \right)$ is in the Lie algebra thereof. Thus, the one-step map for $(\psi_{R}, \psi_{I})$ is both symplectic and unitary \cite{chen2017canonical}. 
 
Finally, the map $\eta_{qm}: \left(\psi,\textbf{A},\textbf{Y}\right)^{n} \mapsto \left(\psi,\textbf{A},\textbf{Y}\right)^{n+1}$ is given by

\begin{equation}
\eta_{qm}: \begin{cases}
        \textbf{A}^{n+1}  = \textbf{A}^{n} , \\
        \textbf{Y}^{n+1}  = \textbf{Y}^{n} + \Delta t \mathcal{J} \left(\textbf{A}^{n},\frac{\psi_{R}^{n+1}+\psi_{R}^{n}}{2},\frac{\psi_{I}^{n+1}+\psi_{I}^{n}}{2} \right) ,\\
    \begin{pmatrix}
        \psi_{R} \\ \psi_{I}
    \end{pmatrix}^{n+1} = \text{Cay}\left[\Gamma \left( \textbf{A}^{n} \right)\right] \begin{pmatrix}
        \psi_{R} \\ \psi_{I}
    \end{pmatrix}^{n}. \\
\end{cases}
\end{equation}

Combining the two solution maps $\eta_{qm}$ and $\eta_{em}$ will allow us to numerically solve the SM system. The matrices $\Lambda$ and $\Gamma\left( \textbf{A}^{n}\right)$ are sparse, meaning that the Cayley transform can be implemented efficiently via any given fast matrix inversion scheme. 

Additionally, the two solution maps can be combined in a variety of ways depending on the desired order for the algorithm. For example, a first-order algorithm with step-size $\Delta t$ is given by 
\begin{equation}\label{mapEquation}
    \eta\left( \Delta t \right)  =  \eta_{qm}\left( \Delta t \right) \circ \eta_{em}\left( \Delta t \right).
\end{equation}
Because both $\eta_{qm}$ and $\eta_{em}$ are symplectic and unitary, so is $\eta$.  A second-order symplectic and unitary method can be constructed as 
\begin{equation}\label{M2}
    \eta^2\left( \Delta t \right)  =  \eta_{qm}\left( \Delta t/2 \right) \circ \eta_{em}\left( \Delta t \right)\circ \eta_{qm}\left( \Delta t/2 \right) .
    \end{equation}
High-order symplectic and unitary methods can be similarly constructed from $\eta^2\left( \Delta t \right) $  as follows \cite{hairer2006geometric,he2015Hamiltonian,chen2017canonical}. For a $2n$-th order method $\eta^{2n}$, a $(2n+2)$-th order method is    
\begin{align}
        \eta^{2(n+1)}\left( \Delta t \right)  =&  \eta^{2n}\left( \alpha\Delta t \right) \circ \eta^{2n}\left( \beta\Delta t \right)\circ \eta^{2n}\left( \alpha\Delta t \right) ,\\
        \alpha=&\frac{1}{2-2^{1/(2n+1)}},\quad\beta=1-2\alpha .
\end{align}

\section{Self-Consistent Electron Radiation Reaction Simulation via SPHINX} \label{simSec}

The algorithms derived in section \ref{algoSection} have been implemented into our MATLAB-based \textbf{S}tructure-\textbf{P}reserving Scr\textbf{h}\"od\textbf{in}ger-Ma\textbf{x}well (\textbf{SPHINX}) solver; we also have a Python port of SPHINX. To implement the Cayley transform we utilize the biconjugate gradient stabilized (bicgstab) method; this, in combination with a completely vectorized integration of our algorithms, allows SPHINX to operate with a high degree of speed, efficiency, and accuracy. In all presented simulations, we utilize an iteration accuracy of $10^{-8}$ for the bicgstab method. SPHINX  currently supports fixed and periodic boundary conditions, and will generate the appropriate discrete representation of the differential operators outlined in Eqs.\,\eqref{divScheme} - \eqref{laplaceScheme} in 1-, 2-, or 3-D. An added benefit of the symplectic splitting scheme we utilize is that by simply bypassing $\eta_{qm}$ or $\eta_{em}$ we can operate the code as either a pure Sch\"ordinger or Maxwell solver respectively without losing any of the conservation properties highlighted in section \ref{algoSection}. 

As a first case study for our code, we leverage SPHINX to solve the SM system for an electron in a uniform magnetic field. These simulations constitute the first time that the fully-coupled nonlinear physics admitted by the SM system has been probed in the context of RR. We can model the classical motion of electron through a quantum coherent state, $\Psi$, given by the following Gaussian wave packet:

\begin{equation} \label{coherent_xy}
    \Psi(x,y) = N_{0} ~\text{exp} \left[-\frac{1}{4\delta^2} |\textbf{r} - \textbf{r}_0|^{2} + i\textbf{k} \cdot \textbf{r} - i\varphi_{0} \right],
\end{equation}

where, $\textbf{r} \equiv \langle x,y \rangle$ is a position vector, $\textbf{r}_0 \equiv \langle x_0,y_0 \rangle$ is the center of the wave packet, $\textbf{k} = \langle k_x, k_y \rangle$ is the wave vector, $ \varphi_{0}$ is a constant phase factor, $\delta = \sqrt{\hbar / q B_{0}^{\text{u}}}$ is the magnetic length defined in terms of the initial uniform magnetic field strength $B_{0}^{\text{u}}$, and $N_0$ is the wavefunction normalization factor. The full details of the derivation of $\Psi$ can be found in Appendix \ref{llAndCsTheoryAppendix}. 


In the present work we make use of atomic units ($\hbar=m_e=|e|=4\pi \epsilon_{0}=1$), however SPHINX was developed to readily adopt any choice of units provided by the user. In what follows, we present 2-D simulations of the dynamics of $\left(\Psi,\textbf{A},\textbf{Y}\right)$ in a uniform magnetic field; this is due in part to the physics of interest being only 2-D, but also due in part to the high spatial resolution required to accurately resolve the dynamics of the coherent state $\Psi$. Spatially our code is normalized with respect to the Bohr radius $a_0 \equiv 4 \pi \epsilon_0 \hbar^2 / |e|^2 m_e$, and we normalize time with respect to the Hartree time $\tau = \hbar / E_H$ where $E_{H} \equiv \hbar^2 / m_e a_0^2$ is the Hartree energy. Unless otherwise stated, our simulations are run over a $[-4a_0,4a_0] \times [-4a_0,4a_0]$ domain with a spatial step size of $\Delta x / a_0 = \Delta y / a_0 = 3.2 \times 10^{-2}$ a temporal step size of $\Delta t / \tau= 1.256 \times 10^{-4}$.


In all simulations presented here, we initialize $(\textbf{A},\textbf{Y})$ as
\begin{equation}
    \begin{split}
        A_{x}^0 / A_0 = & -\frac{\beta}{2}\frac{y}{a_0}, \\
        A_{y}^0 / A_0 = & ~\frac{\beta}{2}\frac{x}{a_0} , \\
        {\textbf{Y}}^0 / Y_0 = & ~ 0,
    \end{split}
\end{equation}
where $A_0$ and $Y_0$ are the field reference magnitudes used to normalize $\textbf{A}$ and $\textbf{Y}$, and $\beta \equiv B_{0}^{\text{u}} / B_0$ is the normalized magnitude of the initial uniform magnetic field; n.b., $B_{0}^{\text{u}}$ is the magnitude of the uniform magnetic field at the start of the simulation and $B_0$ is the magnetic field strength reference value in atomic units. We utilize fixed boundary conditions for the electromagnetic fields such that $B_{z} / B_{0} = \beta$ and $\textbf{Y} / Y_{0} = 0$ on the simulation boundaries. In the case of $(\psi_R, \psi_I)$, we place the guiding center at the origin and set the initial position of the wave packet at $(x_0,y_0) = (1.5a_0, 0)$ and utilize periodic boundary conditions. With the guiding center at the origin, the constant phase $\varphi_0$  vanishes (c.f., Eq.\,\eqref{constantDefs} in appendix \ref{csSection}). As such $(\psi_R, \psi_I)$ are initialized as:
\begin{equation}
    \begin{split}
        {\psi}_{R}^0 / \psi_{0} = & ~\frac{1}{\sqrt{\pi}} e^{-\frac{1}{4 \delta^2} |x-x_0|^2 -\frac{1}{4 \delta^2} y^2} \text{Re} \left\{e^{ik_x x + ik_y y}\right\} ,\\
        {\psi}_{I}^0/ \psi_{0} = & ~ \frac{1}{\sqrt{\pi}} e^{-\frac{1}{4 \delta^2} |x-x_0|^2 -\frac{1}{4 \delta^2} y^2}\text{Im} \left\{e^{ik_x x + ik_y y}\right\} , 
    \end{split}
\end{equation}
where $\psi_0$ is the reference magnitudes used to normalize (\textit{qua} to make unitless) $\psi$ and the $\sqrt{2}$ prefactor is used to be consistent with the definition of $\psi = (\psi_R +i \psi_I) / \sqrt{2}$ outlined in section \ref{contDynamSec}. 

To alleviate the strong separation between the timescales of the coherent state (Schr\"odinger) dynamics and the timescales of the radiative (Maxwell) dynamics -- and ultimately for simulation stability at an affordable time-step size  -- we utilize a reduced speed of light $c = 10^{-2}$ in atomic units. Physically, this reduced speed of light corresponds to an enhancement of quantum electro-dynamical effects. In addition, we examine radiation reaction in electron dynamics at atomic scales. To do so, we choose the magnetic field $B_{0}^{\text{u}}$ such that the magnetic length $\delta$ is comparable to the Bohr radius. Admittedly, achieving this requires an extremely large $B_{0}^{\text{u}}$.  We adopt this choice to reduce computational complexity and isolate the essential physics. Our goal here is to first establish a qualitative physical picture using relatively inexpensive simulations. In future work, we will perform multiscale radiation-reaction simulations with more realistic parameters using HPC facilities. 

\begin{figure}[b!] 
    \centering
    \includegraphics[width=0.9\linewidth]{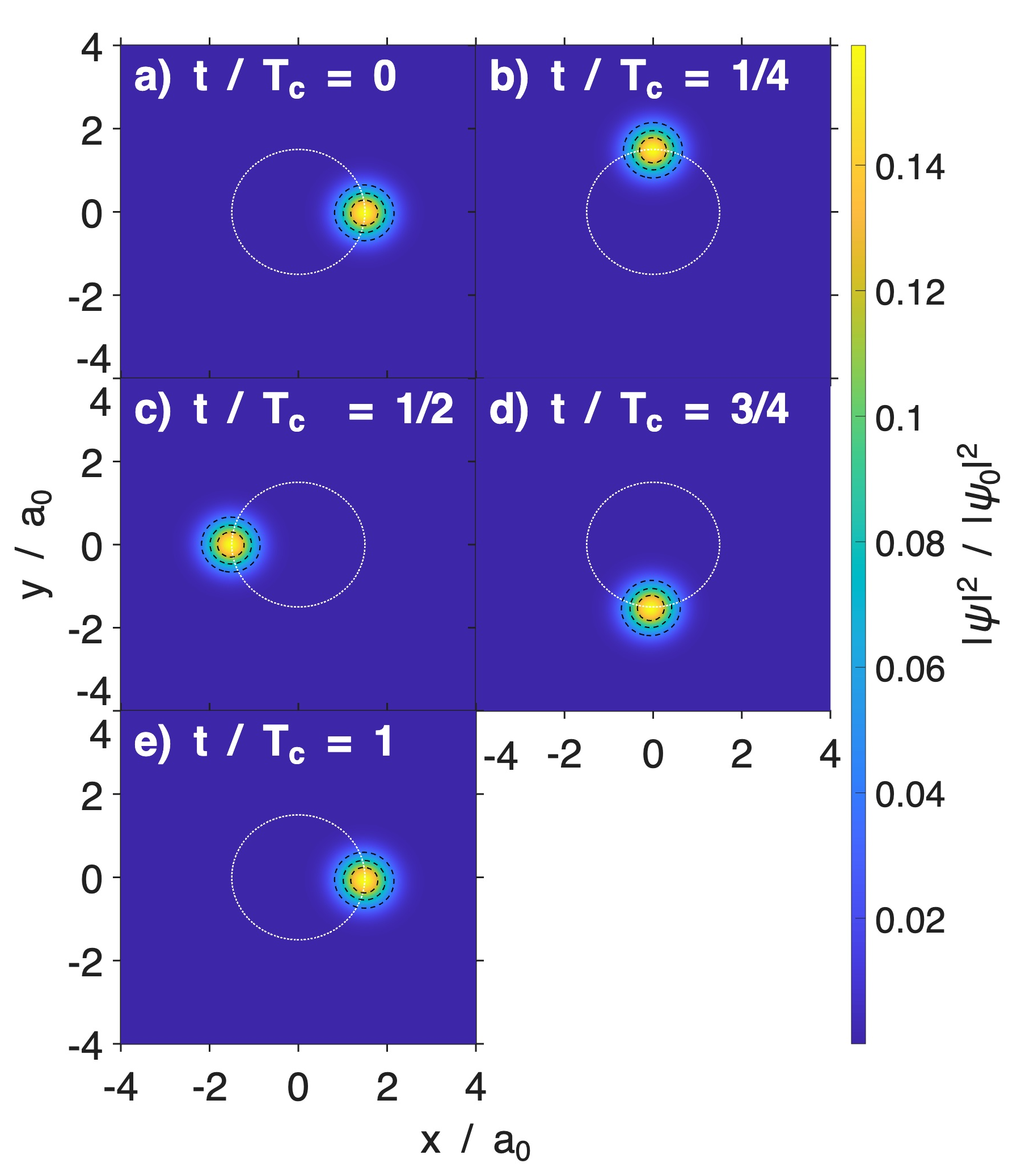}
    \caption{Uncoupled coherent state evolution. Here, $T_{c}  = 2 \pi / \omega_c$ is the cyclotron period and figures (a) - (e) depict the evolution of the coherent state over a full cyclotron period respectively}
    \label{uncoupCyclo}
\end{figure}
\subsection{Coherent State Simulations} \label{csSimulationSection}
\subsubsection{Static Magnetic Field Simulations} \label{sec:staticSims}
To establish a baseline and benchmark SPHINX, we first present the evolution of the coherent state without coupling the Schrodinger and Maxwell systems. Under the influence of a static uniform magnetic field $\Psi$ should evolve as the individual eigenmodes $\psi_{n,m}$ of $\mathcal{H}$ evolve. In section \ref{csSection}, we show that the time-dependent coherent state, $\Psi(t)$, for such a system is given by 

\begin{equation}
    \begin{split}
        \Psi(t) =~& e^{-i\omega_{c}t/2} \Psi|_{w_0 = w_0(t)},\\
        w_0(t) =~& w_0 e^{-i\omega_c t}.
    \end{split}
\end{equation}
That is, for a static uniform magnetic field $\Psi(t)$ is simply the initial $\Psi$ evaluated at $w_0 = w_0 e^{-i\omega_c t}$ and multiplied by a complex oscillating phase factor, where $w_0 = x_0 + i y_0$ and $\omega_c \equiv |q|B_0 / \mu$ is the cyclotron frequency for a particle with charge $q$ and mass $\mu$. All of the time dependence being contained within the constant $w_0$ means that the initial $|\Psi|^2$ is simply advected in time about the fixed guiding center at a radius $\rho_{\ell}$ with frequency $\omega_c$, as expected - c.f., Appendix \ref{csSection} for all details.

\begin{figure}[h!] 
    \centering
    \includegraphics[width=\linewidth]{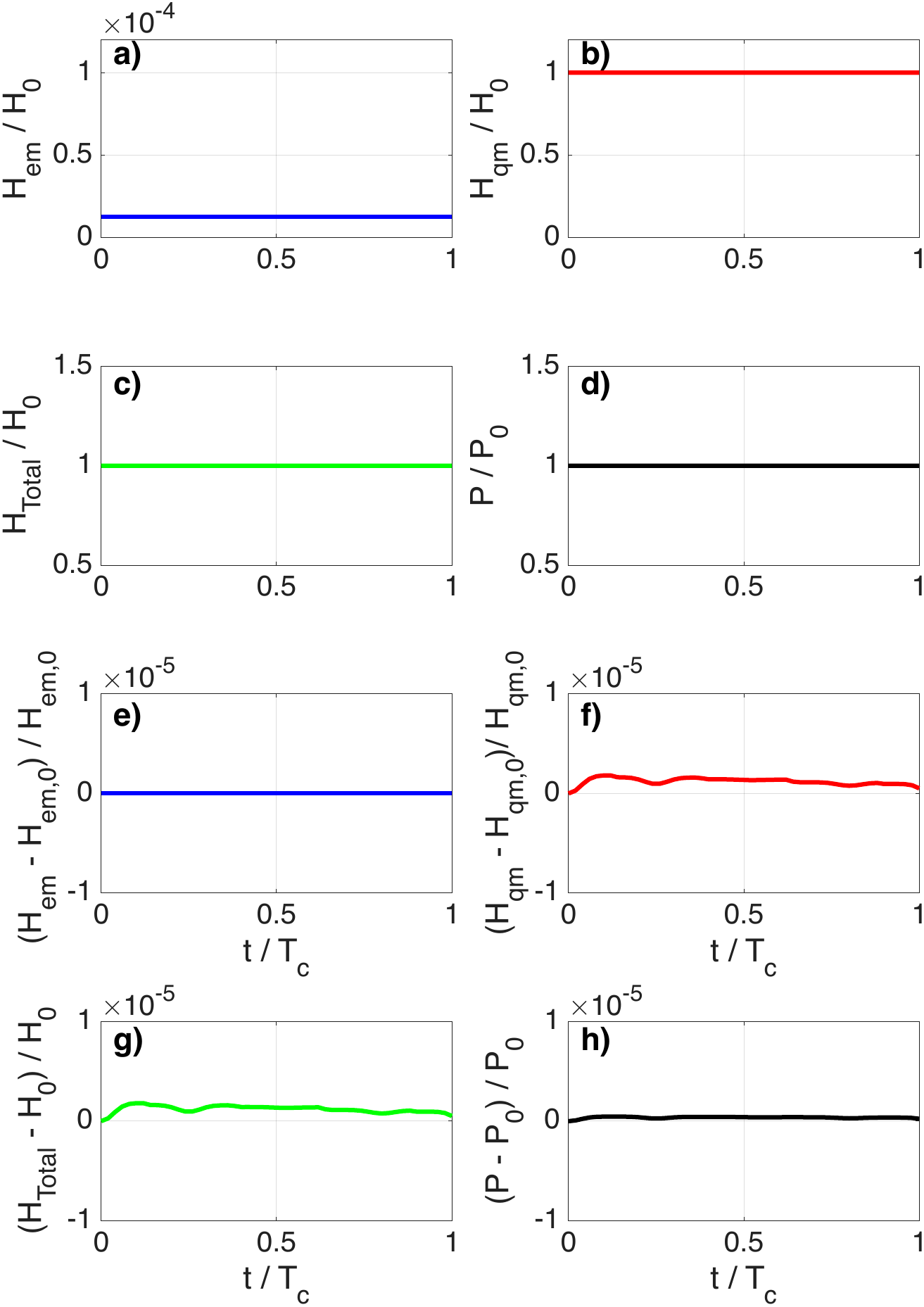}
    \caption{Energy partition evolution of static-field cyclotron simulation. Here, $T_{c}  = 2 \pi / \omega_c$ is the cyclotron period, $H_{\text{Total}} = H_{qm} + H_{em}$ is the total energy of the system, $H_0$ is the initial total energy of the system, $P = \int dV |\psi|^2$, and $P_0$ is the initial value of $P$. Figures (a) - (d) depict evolution of $H_{em}, H_{qm}, H_{\text{Total}}, P$, while figures (e) - (h) respectively depict the error in these values.}
    \label{uncoupPerform}
\end{figure}

In this static field case we take $\beta = 10$ for the uniform magnetic field strength, and simulate our dynamics. In Fig.\,\ref{uncoupCyclo} we present the simulated evolution of $|\Psi|^2$, where we have overlaid a circle of radius $\rho_{\ell}$ atop the simulation as well as the contours of $|\Psi|^2$ for visual aid. Our simulations agree well with the predicted static-field theory, as we see $|\Psi|^2$ follow the expected path for an electron undergoing cyclotron motion in a uniform magnetic field, completing one period in $T_c = \frac{2 \pi}{\omega_c}$. In Fig.\,\ref{uncoupPerform} we present the energy partition and evolution over time. As seen in Figs.\,\ref{uncoupPerform}.a and \ref{uncoupPerform}.b the energy of the system is mostly contained within the quantum subsystem, and the energies of both the quantum and electromagnetic subsystems are independently conserved because the two systems are uncoupled and the electromagnetic fields are held static. As seen in Figs.\,\ref{uncoupPerform}.g and \ref{uncoupPerform}.h, the error in the energy and total probability conservation are bounded by a small amount.

\subsubsection{Coupled Schr\"odinger-Maxwell Simulations} \label{sec:coupledSims}

With a baseline established, we now simulate the fully non-linear dynamics of the coupled Schr\"odinger-Maxwell system. In our algorithms, this constitutes a self-consistent loop in which $\textbf{A}$ and $\textbf{Y}$ evolve due to each other under Maxwell's equations, $\psi_R$ and $\psi_I$ evolve due to each other and $\textbf{A}$ under the Schr\"odinger equation, and $\textbf{Y}$ evolves due to the current $\mathcal{J}\left(\psi,\textbf{A}\right)$. 

\begin{figure}[b!]
    \centering
    \includegraphics[width=\linewidth]{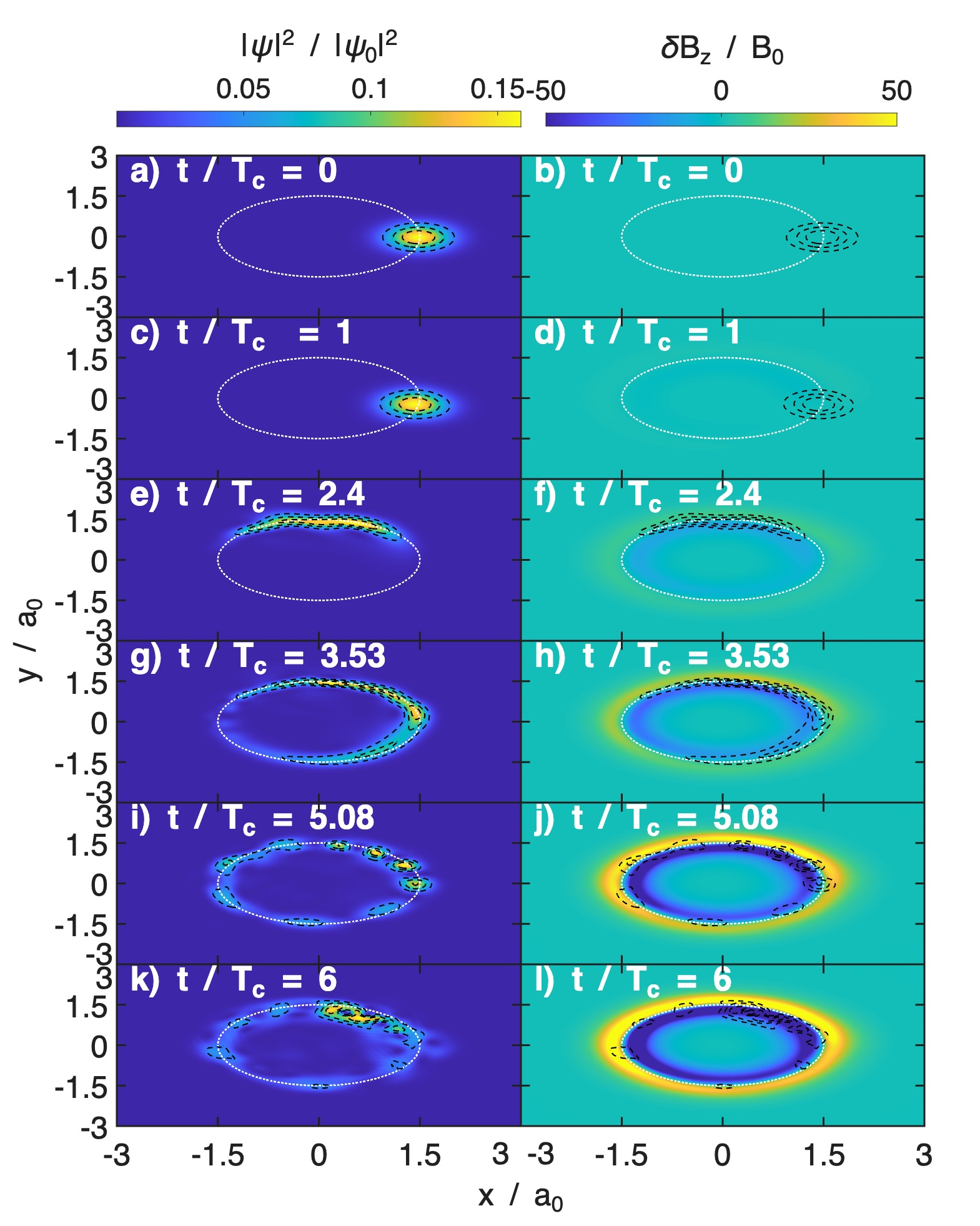}
    \caption{Dynamic evolution of $|\Psi|^2$ and $\delta B_{z}$ over 6 cyclotron periods ($T_{c} = 2 \pi /\omega_c$) for the fully coupled $\beta = 10$ simulation. The left hand column presents the evolution of $|\Psi|^2$ for the coherent state $\Psi$ over the time periods $t /T_{c} = 0,~1,~2.4,~3.53,~5.08, ~\text{and } 6$ in plots (a), (c), (e), (g), (i), and (k) respectively. Similarly, the right hand column presents the evolution of the z-component of the non-uniform/perturbed magnetic field $\delta \textbf{B}$ over the same respective times in plots (b), (d), (f), (h), (j), and (l). In both sets of figures, the contours of $|\Psi|^2$ and a circle of radius $\rho_{\ell}$ (the Larmor radius) are overlaid atop the figures to act as visual aid}
    \label{coupSim10}
\end{figure}
In what follows, we present the dynamic evolution of $|\Psi|^2$ and the z-component of the perturbed magnetic field $\delta B_z / B_0 \equiv B_z/ B_0  - \beta$ for the cases of $\beta = 10$ and $\beta = 5$. To begin, the evolution of $|\Psi|^2$  and $\delta B_z$ in the case of the $\beta = 10$ case can be found in Fig.\,\ref{coupSim10}. Here we see that a single gyro-orbit is completed with minimal distortion to the shape of $|\Psi|^2$, however even after a single orbit we already observe a departure from the ideal cyclotron period (Fig.\,\ref{coupSim10}.c), and as the wave packet completes its gyro-orbit we observe the development of $|\delta B_{z} / B_{0}| \sim 1$ atop the uniform background magnetic field (Fig.\,\ref{coupSim10}.d). As the wave packet completes its second orbit and moving into the third --  sampling the newly established $\delta B_z$ -- we observe significant distortion/elongation to $|\Psi|^2$ (Fig.\,\ref{coupSim10}.e) and an increase in $\delta B_{z}$ to $|\delta B_{z} / B_{0}| \sim 10$ (Fig.\,\ref{coupSim10}.f). The $t / T_{c} = 3.53$ time point ($|\delta B_{z} / B_{0}| \sim 20$)  marks the precipice of the wave packet elongation, after which we begin to observe the fractionation of $|\Psi|^2$ into smaller wave packets along the radius $\rho_{\ell}$ (Fig.\,\ref{coupSim10}.g), and by $t / T_{c} = 5.08$ ($|\delta B_{z} / B_{0}| \sim 60$) we observe that $|\Psi|^2$ has completely fractionated into individual wave packets localized along the Larmor radius $\rho_{\ell}$. This island chain structure is only transiently stable, as by the end of the simulation at $t / T_c = 6$  (Fig \ref{coupSim10}.k) we observe that all coherent structure in $|\Psi|^2$ has been lost and $\delta B_{z}$ reaches a maximum of $|\delta B_{z} / B_{0}| \sim 80$.

\begin{figure}[b!]
    \centering
    \includegraphics[width=\linewidth]{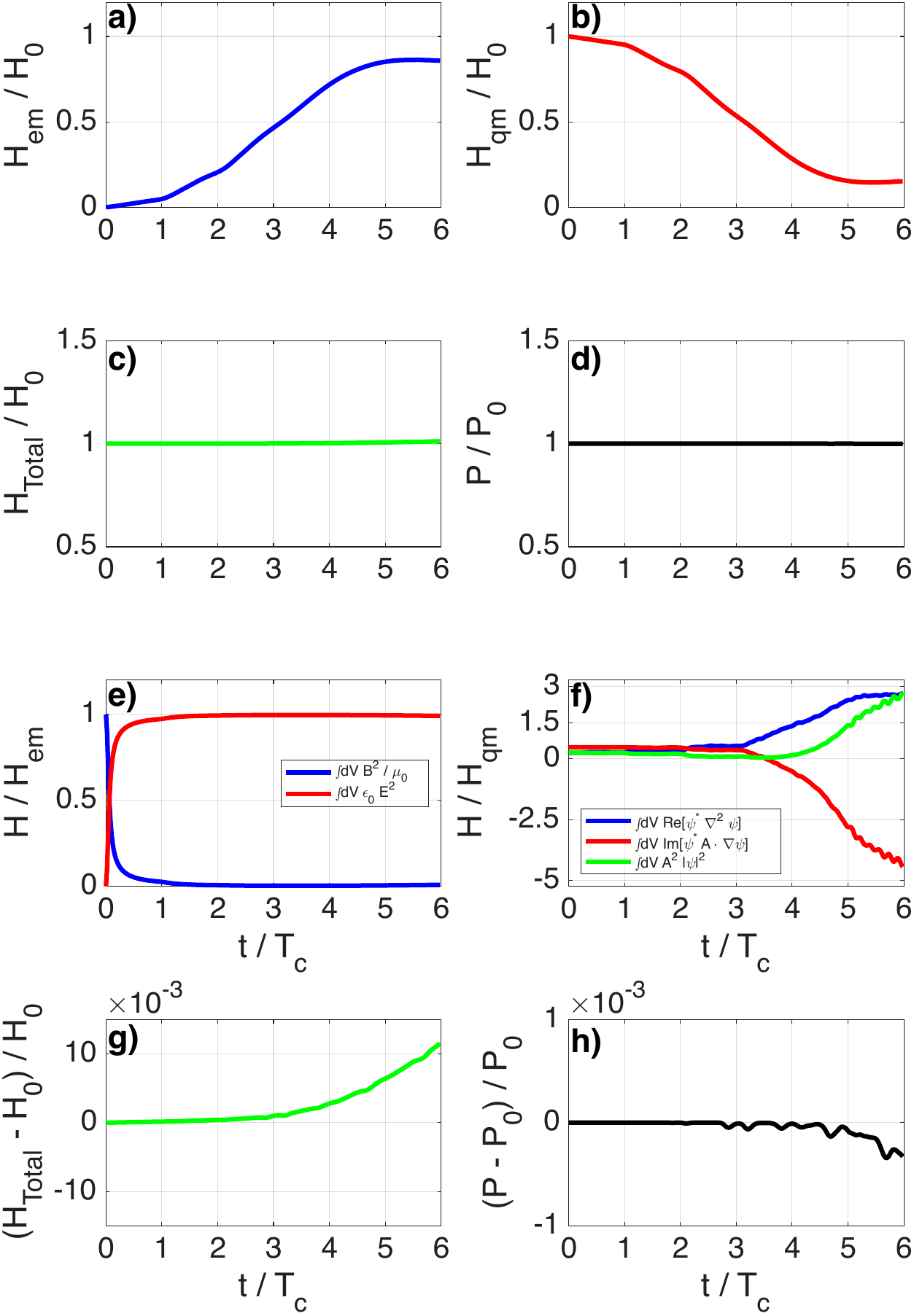}
    \caption{Evolution of the energy partition of the SM system for the fully coupled $\beta = 10$ simulation  over 6 ideal cyclotron periods ($T_{c} = 2 \pi / \omega_c$). Here, $H_{\text{Total}} = H_{qm} + H_{em}$ is the total energy of the system, $H_0$ is the initial total energy of the system, $P \equiv \int dV |\psi|^2$, and $P_0$ is the initial value of $P$. Plots (a)-(c) depict the conservation of the electromagnetic subsystem energy ($H_{em}$), quantum subsystem energy ($H_{qm}$), and $H_{\text{Total}}$ relative to $H_{0}$ respectively, and plot (d) depicts the conservation of $P$ relative to $P_0$. Plots (g) and (h) depicts the error in $H_{\text{Total}}$ and $P$ respectively. Plot (e) and (f) depict how the energy of the partition of each subsystem system evolves relative to the time-dependent subsystem energies $H_{em}(t)$ and $H_{qm}(t)$ respectively}
    \label{coupPerfo10}
\end{figure}

The evolution of the energy partition and total probability throughout this simulation can be found in Fig.\,\ref{coupPerfo10}. Subplots (a), (b), (c), (d), (g), and (h) plot the same data as their respective subplots in Fig.\,\ref{uncoupPerform} except for the fully-coupled $\beta = 10$ simulation in this case. The $\beta = 10$ simulation is run for $6~T_c$ instead of just a single period as in the static field case, therefore the error in the energy conservation is expectedly larger. However, the error in $H_{\text{Total}}$ is still well behaved with the maximum error being $\sim 1 \%$. At the outset of the simulation the energy of the system is mostly stored in the quantum subsystem, and as soon as the simulation begins we observe an energy transfer from the quantum subsystem (Fig.\,\ref{coupPerfo10}.b) to the electromagnetic subsystem (Fig.\,\ref{coupPerfo10}.a). Between $t / T_c = 0$ and $t / T_c = 4$, the radiated power remains approximately constant over each cyclotron orbit, but increases in strength as the wave packet experiences the enhanced $\delta B_{z}$ left in the wake of the previous orbit. By $t / T_{c} = 3$ half of the energy stored in the quantum subsystem has been radiatively transferred into the electromagnetic fields. In Figs.\,\ref{coupPerfo10}.e and \ref{coupPerfo10}.f we plot the evolution of the different components that comprise $H_{em}$ (Eq.\,\eqref{Hem_ref}) and $H_{qm}$ (Eq.\,\eqref{Hqm_ref}) respectively and how they evolve as an instantaneous fraction of their respective total subsystem energy over time. From Fig.\,\ref{coupPerfo10}.e we see that all of the initial energy of the electromagnetic subsystem is stored in the magnetic fields, but as soon as the radiation process begins the energy stored in the electric fields rapidly dominates the $H_{em}$ energy partition. The energy of the quantum subsystem is comprised of three kinetic energy terms: the (1) canonical term $\left( \sim \psi^* \nabla^2 \psi\right)$, (2) paramagnetic term $\left(\sim  \psi^* A \cdot \nabla \psi\right)$, and (3) the diamagnetic term $\left( \sim |\textbf{A}|^2 |\psi|^2 \right)$. Initially, the majority of the energy in the quantum subsystem is stored as paramagnetic energy with the canonical and diamagnetic kinetic energy terms being equal. Once the wave packet completes its first orbit, we see an abrupt drops in the diamagnetic and paramagnetic energy partitions as the radiation process continues. This trend continues along the wave packet's second orbit, with the canonical kinetic energy term dominating the energy partition leading into the third orbit. The dynamics change considerably for $t / T_c > 3$. At $t/T_c = 3.53$ - the same time at which we see $|\Psi|^2$ begin to fractionate - the paramagnetic energy term becomes negative. After $t / T_c = 3$ the radiated power begins to decrease, with the radiation process effectively stopping by $t / T_c = 5$, at which point the canonical and diamagnetic kinetic energies return to occupying the same fraction of the total energy as they did at $t / T_c = 0$. The final energy partition of the quantum subsystem mirrors the initial energy partition: the majority of the energy is stored as paramagnetic kinetic energy (except now negative) with the canonical and diamagnetic kinetic energies occupying roughly equivalent partitions of the remaining energy of the subsystem. 

\begin{figure}[t!]
    \centering
    \includegraphics[width=\linewidth]{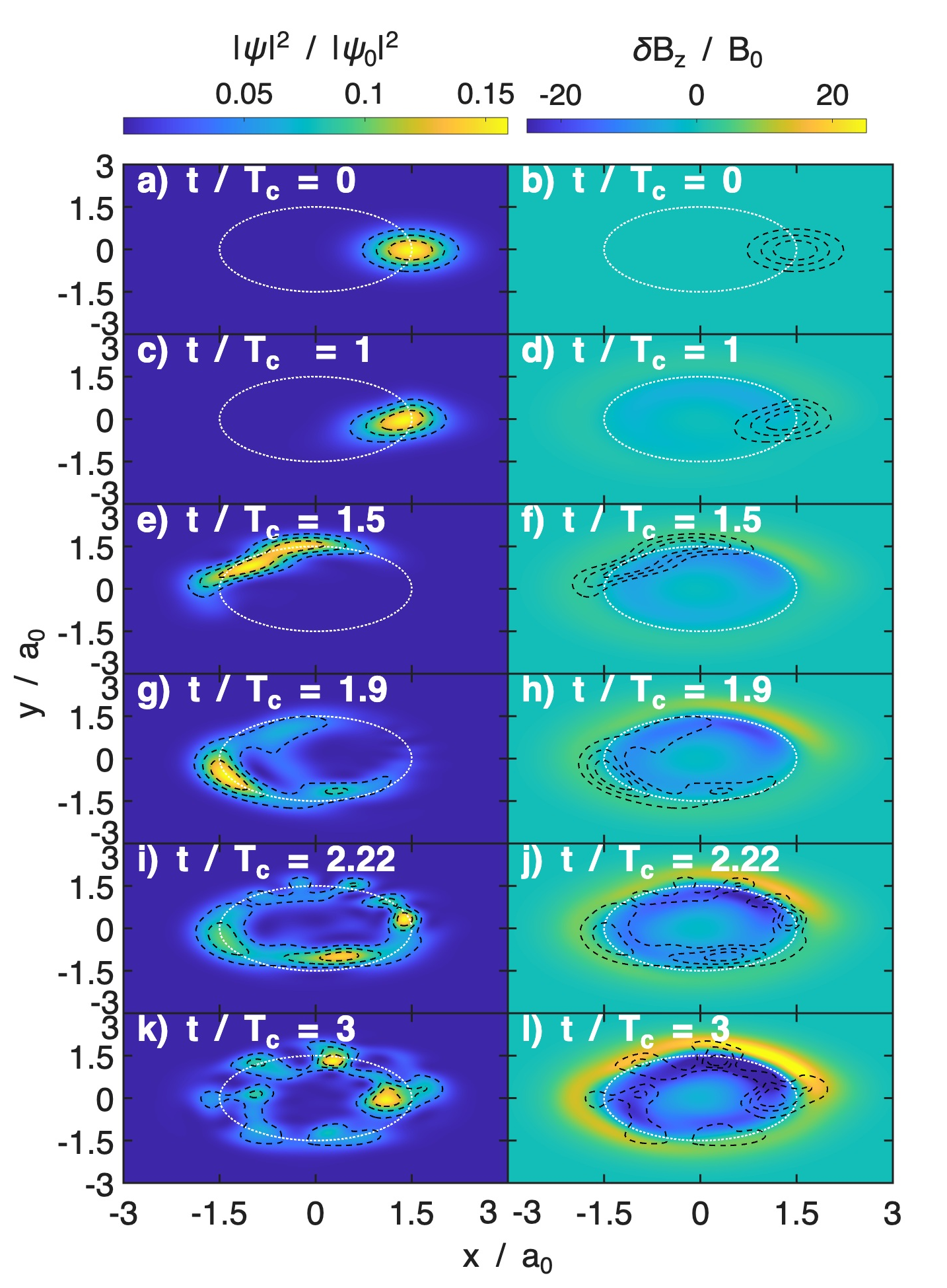}
    \caption{Dynamic evolution of $|\Psi|^2$ and $\delta B_{z}$ over 3 cyclotron periods ($T_{c} = 2 \pi /\omega_c$) for the fully coupled $\beta = 5$ simulation. The left hand column presents the evolution of $|\Psi|^2$ for the coherent state $\Psi$ over the time periods $t /T_{c} = 0,~1,~1.5,~1.9,~2.22, ~\text{and } 3$ in plots (a), (c), (e), (g), (i), and (k) respectively. Similarly, the right hand column presents the evolution of the z-component of the non-uniform/perturbed magnetic field $\delta \textbf{B}$ over the same respective times in plots (b), (d), (f), (h), (j), and (l). In both sets of figures, the contours of $|\Psi|^2$ and a circle of radius $\rho_{\ell}$ (the Larmor radius) are overlaid atop the figures to act as visual aid}
    \label{coupSim5}
\end{figure}

We now turn our attention to the $\beta = 5$ case. It is important to note that all that has changed here is the value of the uniform magnetic field strength at the start of the simulation; i.e., all simulation parameters in the $\beta = 10$ and $\beta = 5$ cases are the same except for the value of the  magnetic field strength at the start. We observe that reducing the value of $\beta$ to $\beta = 5$ increases the radiation reaction process substantially. In Fig.\,\ref{coupSim5}.c we see $|\Psi|^2$ complete its first orbit with significantly more distortion than the $\beta = 10$ case and weaker enhancements to $\delta B_z$ ($|\delta B_z / B_0| \sim 2$). Along the second orbit, we observe significant distortion/elongation  to the wave packet by $t / T_c = 1.5$ (Fig.\,\ref{coupSim10}.e) at which point $\delta B_z$ has increased to $|\delta B_z / B_0| \sim 10$. By the $t / T_c = 1.9$ point the wave packet begins to bifurcate along the ideal cyclotron orbit, with the strength of $\delta B_z$ increasing again to $|\delta B_z / B_0| \sim 15$. This wave-packet bifurcation continues well up to $t / T_c = 2.22$ (Fig.\,\ref{coupSim5}.i), yet we do not observe the same wave-packet fractionation that we observe in the $\beta = 10$ case. By $t / T_c = 3$ (Fig.\,\ref{coupSim5}.k) the wave packet dynamics have completely stopped and we observe a loss of coherent structure in $\Psi$; by the end of the simulation $\delta B_z$ attains a maxima of $|\delta B_z / B_0| \sim 35$. Distinct from the $\beta = 10$ case, the topology/structure of $\delta B_z$ is less altazimuthally symmetric than the $\delta B_z$ established in the $\beta = 10$ case (compare Figs.\,\ref{coupSim10}.i and \ref{coupSim5}.i)

\begin{figure}[b!]
    \centering
    \includegraphics[width=\linewidth]{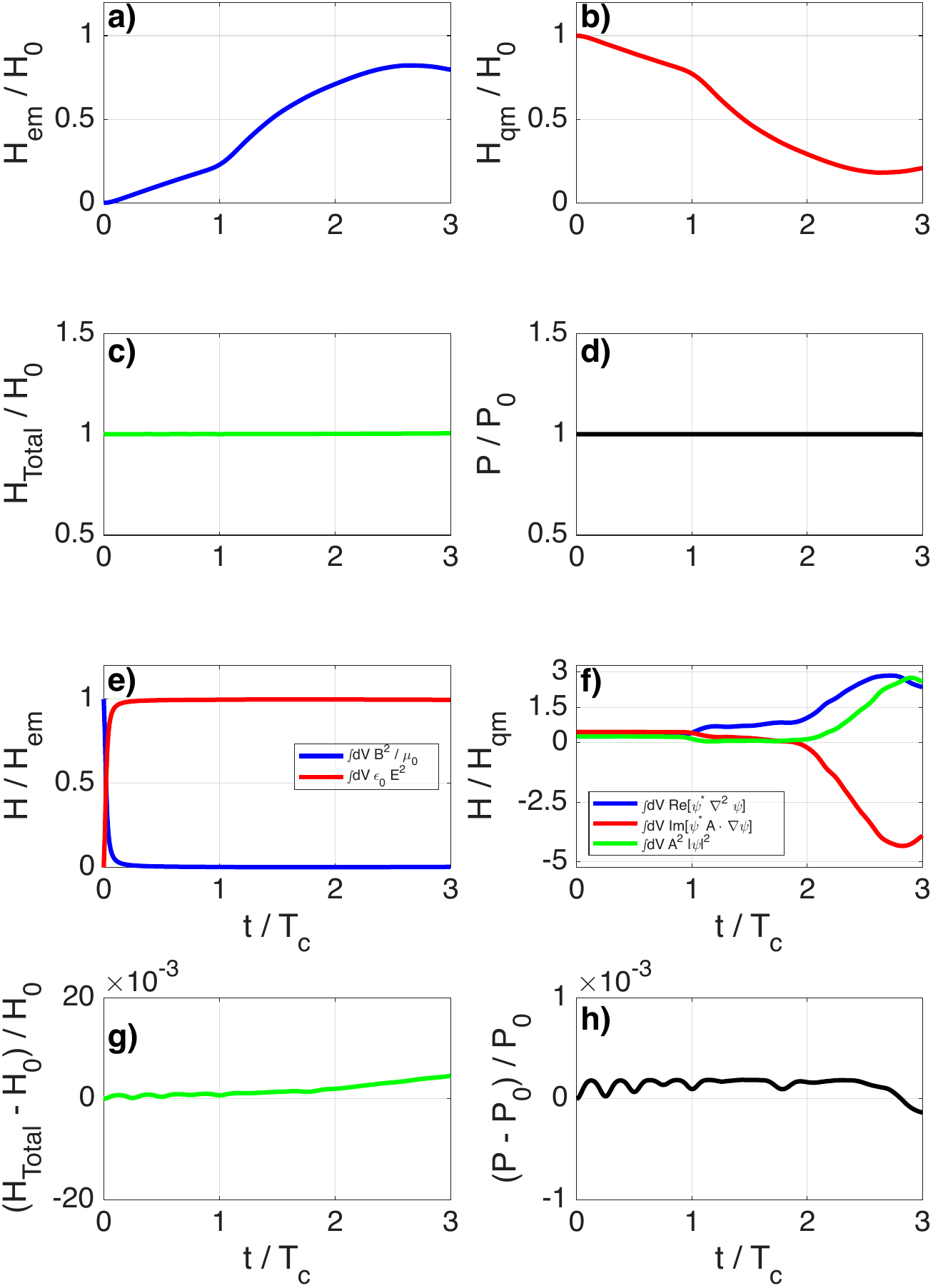}
    \caption{Evolution of the energy partition of the SM system for the fully coupled $\beta = 5$ simulation over 3 ideal cyclotron periods ($T_{c} = 2 \pi / \omega_c$). Here, $H_{\text{Total}} = H_{qm} + H_{em}$ is the total energy of the system, $H_0$ is the initial total energy of the system, $P \equiv \int dV |\psi|^2$, and $P_0$ is the initial value of $P$. Plots (a)-(c) depict the conservation of the electromagnetic subsystem energy ($H_{em}$), quantum subsystem energy ($H_{qm}$), and $H_{\text{Total}}$ relative to $H_{0}$ respectively, and plot (d) depicts the conservation of $P$ relative to $P_0$. Plots (g) and (h) depicts the error in $H_{\text{Total}}$ and $P$ respectively. Plot (e) and (f) depict how the energy of the partition of each subsystem system evolves relative to the time-dependent subsystem energies $H_{em}(t)$ and $H_{qm}(t)$ respectively}
    \label{coupPerfo5}
\end{figure}

In Fig.\,\ref{coupPerfo5} we present the evolution of the energy partition of the $\beta = 5$ case. Just as with the $\beta = 10$ case we observe that as soon as the wave packet begins its cyclotron orbit, energy is radiatively transferred out of the quantum subsystem and into the electromagnetic subsystem (Figs.\,\ref{coupPerfo5}.a and \ref{coupPerfo5}.b). The radiated power remains approximately constant over the first period, and achieves its maximum in between $t / T_c = 1$ and $t / T_c = 2$, after which the radiated power decreases and effectively ceases by $t / T_c = 3$. By $t / T_c = 1.5$, half of the energy of the quantum subsystem has been radiatively transferred to the electromagnetic fields. Qualitatively the evolution of the energy partition of the individual quantum and electromagnetic subsystems (Figs.\,\ref{coupPerfo5}.e and \ref{coupPerfo5}.f) are virtually identical to the $\beta = 10$ case (Figs.\,\ref{coupPerfo10}.e and \ref{coupPerfo10}.f), only faster. Interestingly, in the $\beta = 5$ case the paramagnetic energy becomes negative at approximately $t / T_c \sim 2.06$, yet we don't observe such clear fractionation of the wave packet after this point of time as we did in the $\beta = 10$ case. 

\subsection{Landau Level Simulations} \label{sec:llSims}
\begin{figure}
    \centering
    \includegraphics[width=\linewidth]{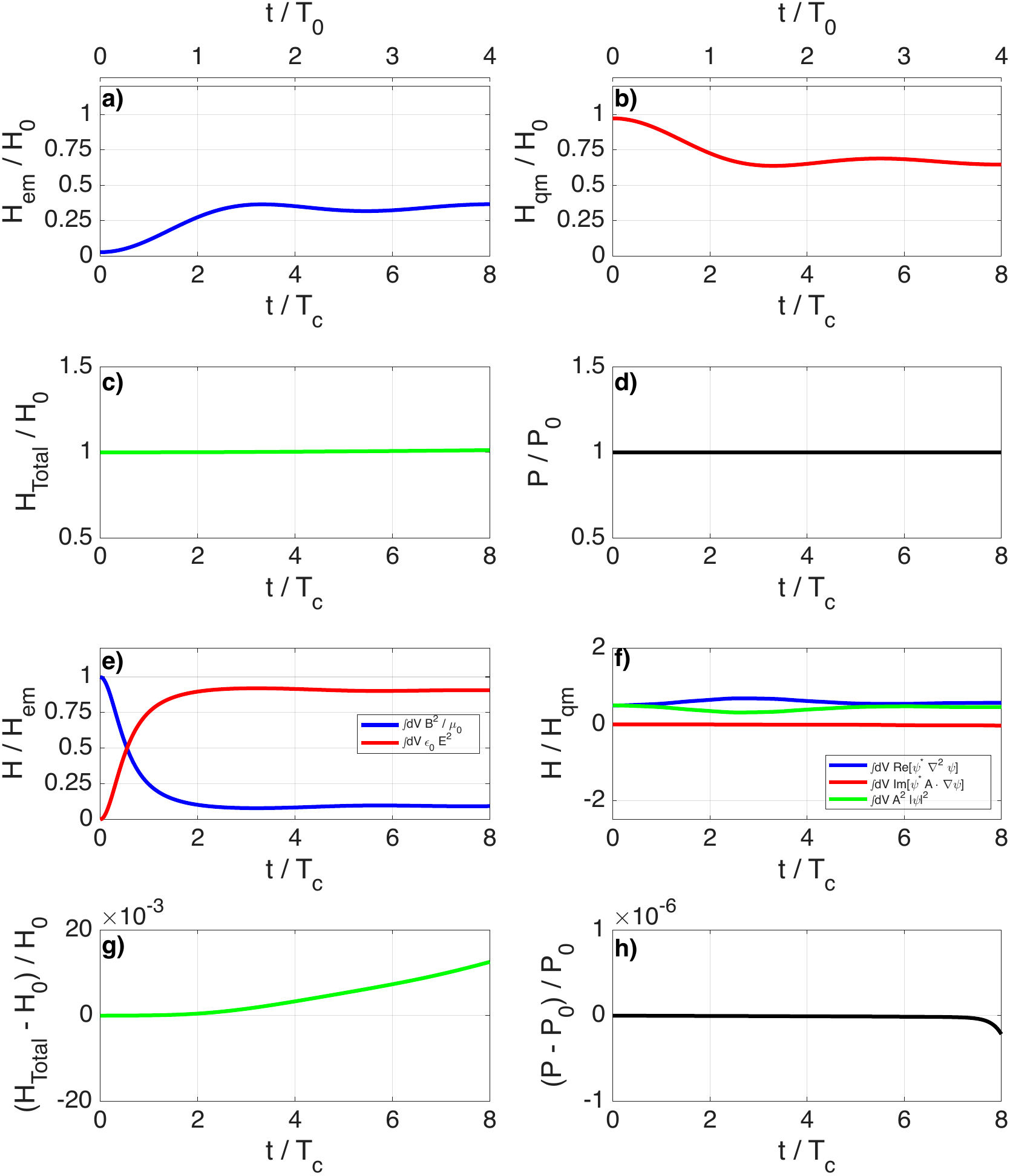}
    \caption{Evolution of the energy partition of the SM system for the (n,m) = (0,0) Landau level eigenmode over two ideal eigenmode periods $T_{0} = 2 \pi / \omega_{0}$, where $\omega_{0} = \frac{1}{2} \omega_c$. Here, $H_{\text{Total}} = H_{qm} + H_{em}$ is the total energy of the system, $H_0$ is the initial total energy of the system, $P \equiv \int dV |\psi|^2$, and $P_0$ is the initial value of $P$. Plots (a)-(c) depict the conservation of the electromagnetic subsystem energy ($H_{em}$), quantum subsystem energy ($H_{qm}$), and $H_{\text{Total}}$ relative to $H_{0}$ respectively, and plot (d) depicts the conservation of $P$ relative to $P_0$. Plots (g) and (h) depicts the error in $H_{\text{Total}}$ and $P$ respectively. Plot (e) and (f) depict how the energy of the partition of each subsystem system evolves relative to the time-dependent subsystem energies $H_{em}(t)$ and $H_{qm}(t)$ respectively}
    \label{fig:n0m0_perfo}
\end{figure}

Let us consider now the individual Landau level eigenstates. If the electromagnetic fields are held-fixed and not self-consistently evolved with the quantum subsystem, then the dynamics of a given Landau level eigenstate, $\psi_{n,m}$, are given by:
\begin{equation}
    \begin{split}
        \psi_{n,m}(t) =& ~\psi_{n,m}(0) e^{-i E_n t / \hbar} ,\\
        E_n =& ~\hbar \omega_c (n+\frac{1}{2}), \\
    \end{split}     
\end{equation}
where $n$ is the Landau level energy index with degeneracy index $m$, and $\psi_{n,m}(0)$ is the initial wavefunction at $t=0$. The dynamic evolution of the Landau levels, however, when considering the fully-coupled SM system is not so trivial. In this section we present the fully-coupled non-linear evolution of the individual non-degenerate ($m=0$) ground ($n = 0$) and first excited ($n = 1$) states of the 2D Landau levels - the full derivation of such states can be found in appendix \ref{LLderiv}. To maintain consistency with the results of section \ref{csSimulationSection} we adopt the same simulation parameters, electromagnetic field initialization, and boundary conditions as in the $\beta = 10$ case, with the exception of normalized step-size: in the $(n,m) = (0,0)$ simulation $\Delta t / \tau = 2.51 \times 10^{-4}$ and in the $(n,m) = (1,0)$simulation we utilize $\Delta t / \tau = 8.376 \times 10^{-5}$. The $(n,m) = (0,0)$ state is initialized as

\begin{equation}
    \begin{split}
        {\psi}_{R}^0 / \psi_{0} = & ~\frac{1}{\sqrt{\pi}}  e^{-\frac{1}{4 \delta^2} (x^2 + y^2)} ,\\
        {\psi}_{I}^0/ \psi_{0} = & ~ 0. 
    \end{split}
\end{equation}
In the ideal (i.e., static field) case, the expected frequency of such a state is $\omega_0 = \omega_c / 2 $ with a period of $T_0 = 2 \pi / \omega_0$. Below in Fig.\,\ref{fig:n0m0} we present the evolution of $\text{Re}\left[\psi\right]$, $\text{Im}\left[\psi\right]$, and $\delta B_z$ simulated over 4 of the ideal ground state eigenmode periods $T_0$, and in Fig.\,\ref{fig:n0m0_perfo} we present the energetic data for this simulation. The evolution of $\psi_{0,0}$ and the radiative process tells a similar story to that observed in section \ref{sec:coupledSims}: the motion of $\psi_{0,0}$ immediately begins a radiative transfer of energy out of the quantum subsystem and into the electromagnetic subsystem, as seen in Fig.\,\ref{fig:n0m0_perfo}.a. As $\psi_{0,0}$ continues to radiate, we observe that $\text{Re}\left[\psi\right]$ and $\text{Im}\left[\psi\right]$ slowly dampen/depart from their ideal periodic motion, with the first period being completed at $t / T_0 = 1.14$ (Figs.\,\ref{fig:n0m0}.g and \ref{fig:n0m0}.h), and the second period being delayed even more substantially ($t / T_0 = 2.26$). Initially, the energy lost by the quantum subsystem in the case of  $\psi_{0,0}$ is comparable to that of the coherent state $\Psi$ until $t / T_c = 3$ (compare Figs.\,\ref{fig:n0m0_perfo}.b and \ref{coupPerfo10}.b). Once $\psi_{0,0}$ stops radiating at around $t / T_c = 3$ (c.f., \ref{fig:n0m0_perfo}.b), $\psi_{0,0}$ remains stable and continues to oscillate at a dampened frequency. Another distinct feature of these simulations compared to full coherent state simulations is the asymmetry of $\delta B_z$. The magnitude of the outer (positive) ring of $\delta B_z$ (e.g., Fig.\,\ref{fig:n0m0}.o) is approximately half of the magnitude of the inner (negative) ring. The maximum value of $\delta B_z$ at the end of the simulation is $\delta B_z / B_0 \sim 14$ and corresponding minimum value is $\delta B_z / B_0 \sim -28$, substantially less than the observed $\delta B_z$ in the coherent state simulations.

\begin{figure}
    \centering
    \includegraphics[width=\linewidth]{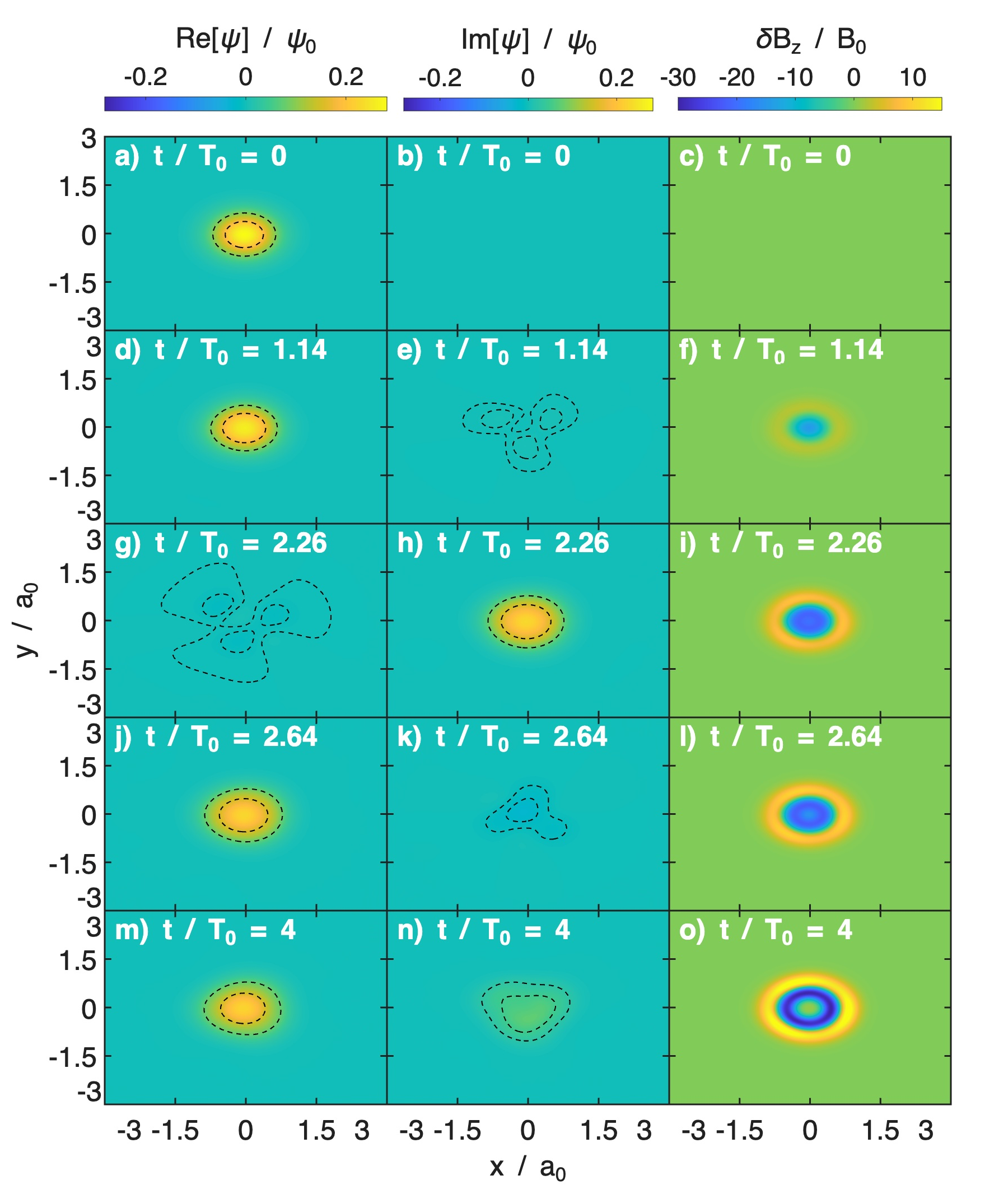}
    \caption{Dynamic evolution of the $\text{Re}\left[\psi_{0,0}\right] / \psi_0$, $\text{Im}\left[\psi_{0,0}\right]  / \psi_0$, and $\delta B_z  / B_0$ over 4 ideal eigenmode periods $T_0 \equiv \frac{2 \pi}{\omega_{0}}$, where $\omega_{0} = \frac{1}{2}\omega_c$. The first column presents snapshots of the evolution of $\text{Re}\left[\psi_{0,0}\right]$ at times $t/T_0 = (a)~0,~(d)~1.14, ~(g)~2.26, ~(j)~2.64, \text{ and} ~(m)~4$. The second column presents the same snapshots at the same times for $\text{Im}\left[\psi_{0,0}\right]$ in subplots (b), (e), (h), (k), and (n) respectively, and the third column again presents the same data for the perturbed magnetic field $\delta B_z$ in in subplots (c), (f), (i), (l), and (o).}
    \label{fig:n0m0}
\end{figure}

\begin{figure}
    \centering
    \includegraphics[width=\linewidth]{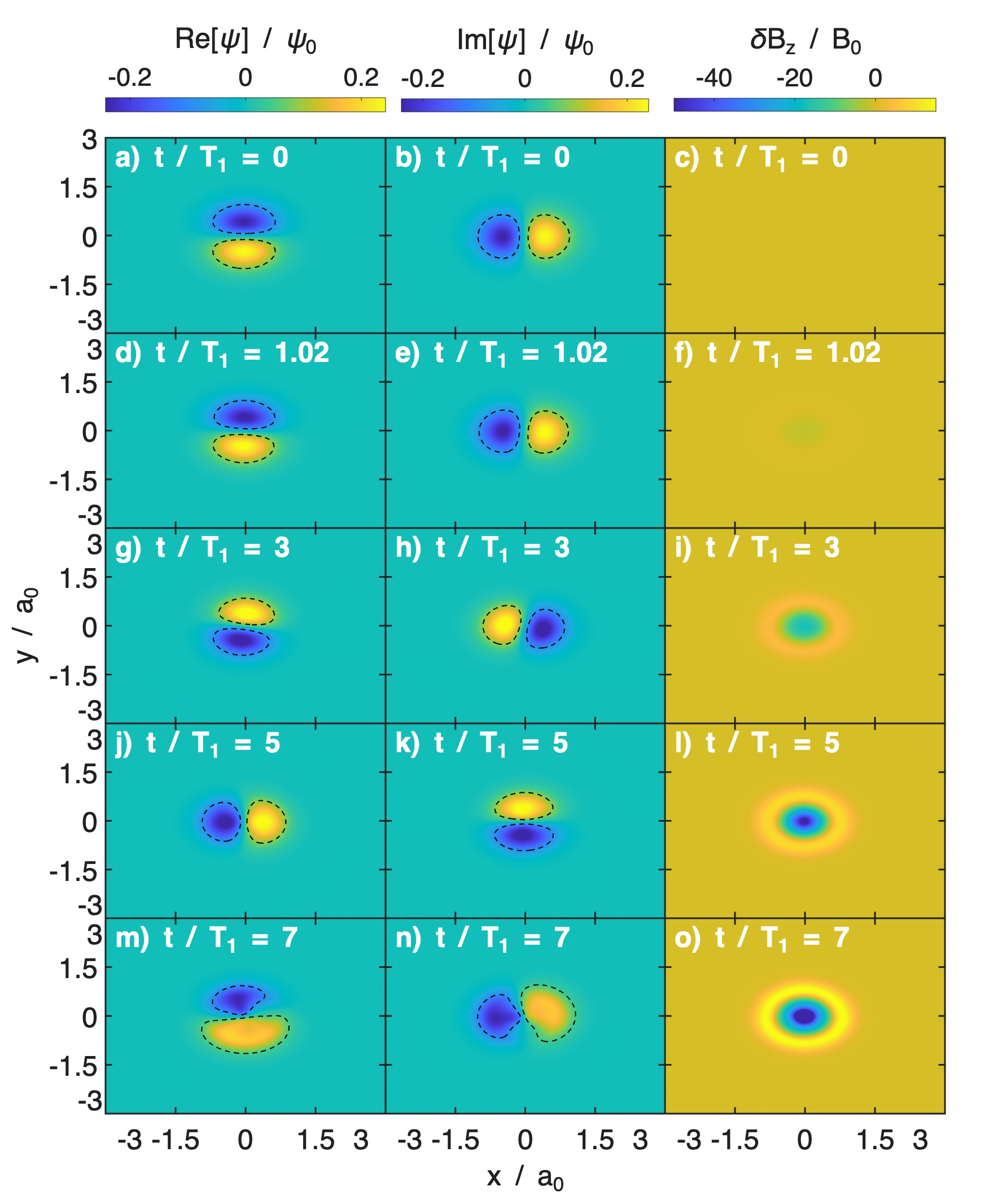}
    \caption{Dynamic evolution of the $\text{Re}\left[\psi_{1,0}\right] / \psi_0$, $\text{Im}\left[\psi_{1,0}\right]  / \psi_0$, and $\delta B_z  / B_0$ over 7 ideal eigenmode periods $T_1 \equiv \frac{2 \pi}{\omega_1}$, where $\omega_1 = \frac{3}{2}\omega_c$. The first column presents snapshots of the evolution of $\text{Re}\left[\psi_{1,0}\right]$ at times $t/T_0 = (a)~0,~(d)~1.02, ~(g)~3, ~(j)~5, \text{ and} ~(m)~7$. The second column presents the same snapshots at the same times for $\text{Im}\left[\psi_{1,0}\right]$ in subplots (b), (e), (h), (k), and (n) respectively, and the third column again presents the same data for the perturbed magnetic field $\delta B_z$ in in subplots (c), (f), (i), (l), and (o).}
    \label{fig:n1m0}
\end{figure}

\begin{figure}
    \centering
    \includegraphics[width=\linewidth]{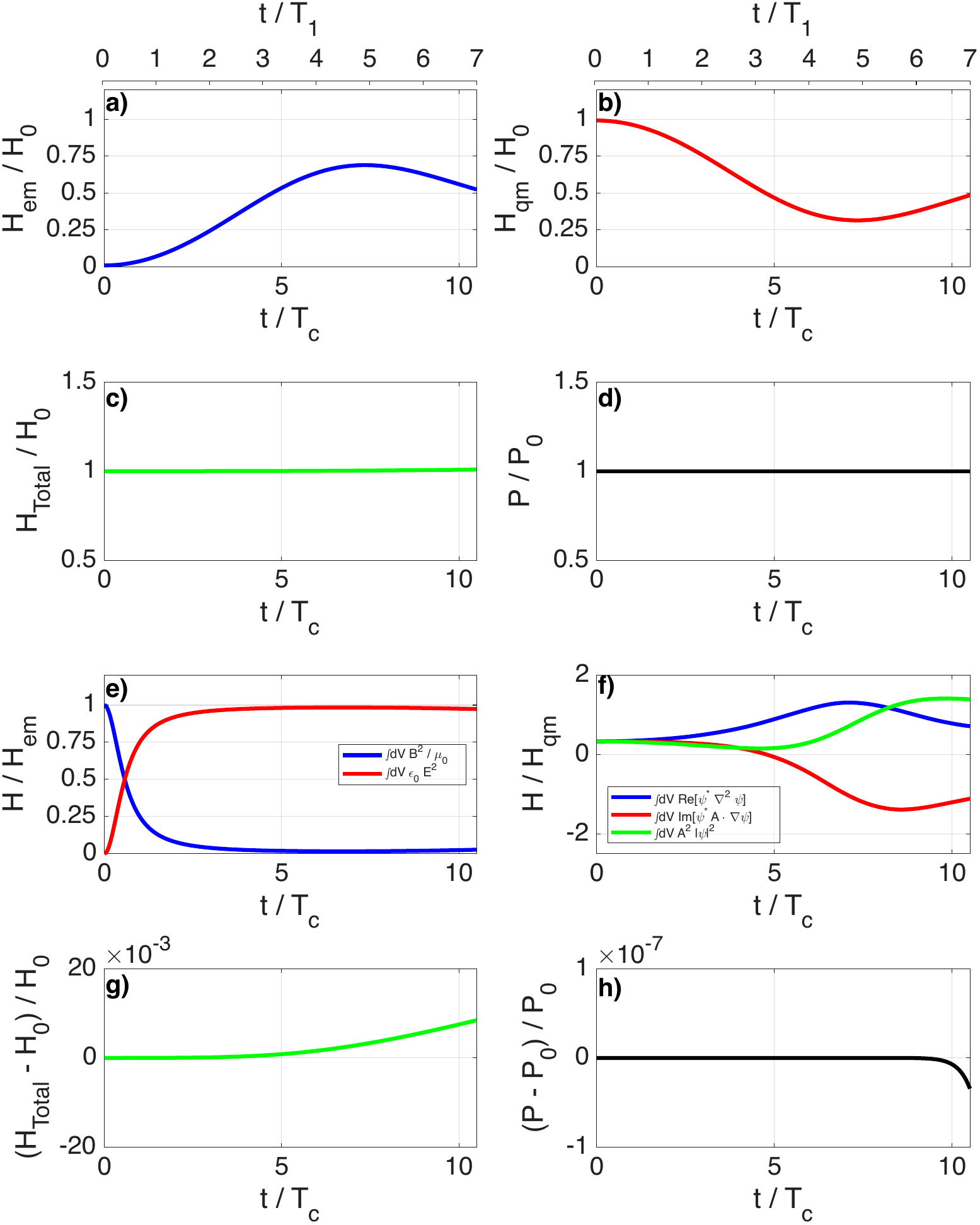}
    \caption{Evolution of the energy partition of the SM system for the (n,m) = (1,0) Landau level eigenmode over two ideal eigenmode periods $T_{1} = 2 \pi / \omega_1$, where $\omega_1 = \frac{3}{2} \omega_c$. Here, $H_{\text{Total}} = H_{qm} + H_{em}$ is the total energy of the system, $H_0$ is the initial total energy of the system, $P \equiv \int dV |\psi|^2$, and $P_0$ is the initial value of $P$. Plots (a)-(c) depict the conservation of the electromagnetic subsystem energy ($H_{em}$), quantum subsystem energy ($H_{qm}$), and $H_{\text{Total}}$ relative to $H_{0}$ respectively, and plot (d) depicts the conservation of $P$ relative to $P_0$. Plots (g) and (h) depicts the error in $H_{\text{Total}}$ and $P$ respectively. Plot (e) and (f) depict how the energy of the partition of each subsystem system evolves relative to the time-dependent subsystem energies $H_{em}(t)$ and $H_{qm}(t)$ respectively}
    \label{fig:n1m0_perfo}
\end{figure}

Finally, the $(n,m) = (1,0)$ excited state is initialized as
\begin{equation}
    \begin{split}
        {\psi}_{R}^0 / \psi_{0} = & ~-\frac{1}{\sqrt{2\pi}}\frac{y}{\delta}  e^{-\frac{1}{4 \delta^2} (x^2 + y^2)} ,\\
        {\psi}_{I}^0/ \psi_{0} = & ~\frac{1}{\sqrt{2\pi}}\frac{x}{ \delta}  e^{-\frac{1}{4 \delta^2} (x^2 + y^2)} , 
    \end{split}
\end{equation}
which is easily obtained via Eq.\,\eqref{psi_nm_adag} in section \ref{LLderiv} of our appendix. The evolution of $\text{Re}\left[\psi\right]$, $\text{Im}\left[\psi\right]$, and $\delta B_z$ for this simulation can be found in Fig.\,\ref{fig:n1m0}, where we present snapshots over 7 ideal eigenmode periods, $T_1 = \frac{2 \pi}{\omega_1}$ where $\omega_1 = \frac{3}{2} \omega_c$. In general, it appears that the most significant difference between the ideal (Schr\"odinger-only) and coupled (full SM) dynamics for the individual eigenmodes is that the energy lost by the quantum subsystem during the radiation process in the coupled case serves to dampen the ideal oscillatory motion. From Fig.\,\ref{fig:n1m0}.d we see that the period of the $\psi_{1,0}$ is slightly delayed, but not by much given that the radiation process is weaker in this case. By $t / T_1 = 3$ (Figs.\,\ref{fig:n1m0}.g and \ref{fig:n1m0}.h) the accumulated effects of the radiation are more readily seen as by the time the third period should be completed, the orbitals are effectively delayed by $t / T_1 = 0.5$. In Fig.\,\ref{fig:n1m0_perfo} we present the same data presented in Fig.\,\ref{fig:n1m0_perfo} except now for the  $(n,m) = (1,0)$ excited state, and from the subplot \ref{fig:n1m0_perfo}.a we see that at $t / T_1 = 3$ approximately half to the energy of the quantum subsystem has been radiatively transferred to the electromagnetic fields. By $t / T_1 = 5$ the radiation process has effectively stopped and we observe a shift in the qualitative dynamics of $\psi_{1,0}$. Heretofore the radiative energy transfer has served to dampen the periodic motion of $\psi_{1,0}$ but not distort the profile of the orbitals themselves; for $t / T_c > 5$ we begin to observe the profile shape of $\psi_R$ and $\psi_I$ deviate from their ideal forms (Figs.\,\ref{fig:n1m0}.m and \ref{fig:n1m0}.n). This qualitative shift in the dynamics of $\psi_{1,0}$ corresponds to a period in which energy is being transferred out of the electromagnetic subsystem (Fig.\,\ref{fig:n1m0_perfo}.a) and back into the quantum subsystem (Fig.\,\ref{fig:n1m0_perfo}.b). Also of note, the asymmetric magnetic perturbation $\delta B_z$ observed in the $(n,m) = (0,0)$ simulation is stronger in the $(n,m) = (1,0)$ simulation with $\delta B_z$ obtaining a maxima of $\delta B_z / B_0 \sim 26$ and a minima of $\delta B_z / B_0 \sim -110$.  

\section{Conclusions} \label{concSec}
We have presented geometric structure-preserving algorithms to for SM systems which we have integrated into our Structure Preserving scHr\"odInger-maXwell (SPHINX) solver. As a pilot study for SPHINX, we present two case studies: (1) two sets of simulations of the motion of a coherent state constructed from the Landau level eigenstates of the Hamiltonian for a (spinless) electron in a uniform magnetic field, one set is a control test with a static magnetic field and the second set are fully-coupled simulations in which the electromagnetic fields and particle dynamics are evolved self consistently for two different initial uniform magnetic field strengths and (2) two simulations of the fully-coupled non-degenerate ($m = 0$) ground ($n=0$) and first-excited ($n=1$) state Landau levels. In case 1 our simulations agree with the analytical theory for the control case of the static magnetic field, as we observe that the probability density $|\Psi|^2$ is advected about the guiding center at a fixed radius (the Larmor radius, $\rho_{\ell}$) in one cyclotron period, $T_c$. Our fully-coupled simulations are run at two different initial uniform magnetic field strengths, $B_z / B_0 = 10$ and $B_z / B_0 = 5$. The $B_z / B_0 = 10$ simulation shows that the quantum coherent state completely decoheres within 6 ideal cyclotron periods, completely undermining the very concept of a RR force. Prior to total decoherence, we observe the single quantum wave packet fractionate into a transiently stable island chain structure of smaller wave packets along the ideal cyclotron orbit. Energetically, this fractionation process exactly coincides with point in time at which the paramagnetic energy $H_{\text{para}} \sim -\frac{q}{m} \text{Im}\left[ \psi^* \textbf{A} \cdot \nabla \psi \right]$ becomes negative. The observed wave-packet fractionation, island-chain formation, and decoherence have yet to be observed in the analytical theory. Our second simulation reduces the initial magnetic field strength in half to $B_z / B_0 = 5$, which resulted in a \textit{faster} RR process in which the coherent state has been totally destroyed by within 3 ideal cyclotron periods. While the $B_z / B_0 = 5$ and $B_z / B_0 = 10$ simulations were energetically similar, they were dynamically very different despite only changing the strength of the magnetic field. Equivalently, it is straight forward to show that the ratio of the magnetic length $\delta$ in the $B_z / B_0 = 5$ case ($\delta_{5}$) and the $B_z / B_0 =10$ case ($\delta_{10}$) is $\delta_{5} / \delta_{10} = \sqrt{2}$, meaning that our stronger RR correlates with larger wave-packet size.

As an additional study, we also presented the self-consistent evolution of the non-degenerate ($m = 0$) ground ($n=0$) and first-excited ($n=1$) Landau level eigenmodes under the same simulation set up as our $B_z / B_0 = 10$ case study. Here, we found that the ground state eigenmode radiation to be somewhat comparable to the full coherent state at the outset, before the radiation process ceased completely and the system relaxed. Simulation of the first-excited ($n=1$) Landau level eigenmode showed that coupling the Schr\"odinger and Maxwell systems generally serves to dampen the motion of the eigenmodes severely as the radiative transfer of energy occurs, however in the $n=1$ simulation we also began to observe distortion to the profiles of $\text{Re}\left[ \psi_{1,0} \right]$ and $\text{Im}\left[ \psi_{1,0} \right]$ once the energy transfer reversed and flowed from the electromagnetic fields back into the quantum subsystem. 


In the context of the classic canon of RR literature, the presented work constitutes a novel view of the RR process from a semi-classical perspective in which the SM system has been self-consistently evolved numerically for a quantum particle in a uniform magnetic field. We find that the full dynamical picture when including non-linearity complicates the physical portrait of RR substantially, and that accounting for particle size and geometry is important. Our work supports the view that the RR problem is best understood as a pathological consequence of the point-particle idealization being used outside its domain of validity, not as a failure of the underlying electrodynamic, classical, and/or quantum theories themselves. The intended goal of this work is to provide a first, computationally inexpensive, view of the nonlinear RR physics admitted by the SM system. Future work could improve upon this work with access to more computation power by relaxing the extreme-field (i.e., large magnitude of the initial uniform magnetic field strength) and reduced speed of light assumptions that made the presented suite of simulations feasible. Additionally, the role of quantum field theory effects in such systems could be better understood via similar studies instead using the Dirac-Maxwell (DM) system of equations.

\section{Acknowledgments}
The simulations presented in this article were performed on computational resources managed and supported by Princeton Research Computing, a consortium of groups including the Princeton Institute for Computational Science and Engineering (PICSciE) and Research Computing at Princeton University.  J. M. Molina acknoweldges support of NSF GRFP (KB0013612).

\appendix
\section{Landau Levels Derivation and Coherent State Construction} \label{llAndCsTheoryAppendix}
In the following section we review the construction of coherent states for a spinless negatively charged particle in a uniform magnetic field $\textbf{B} = B_{0} \textbf{e}_{z}$, where $B_{0}$ is the magnitude of the uniform magnetic field \cite{Zhu2017,Malkin1969,Feldman1970,Kowalski2005}. We begin by first deriving the energy eigenstates for our system---the well known Landau levels. Using these energy eigenstates, we can construct a minimally uncertain Gaussian wave packet (i.e., a coherent state) that capture the dynamics of our classical charged particle.
\subsection{Landau Level Derivation} \label{LLderiv}
 For a particle with mass $\mu$ and charge $\mathcal{Q} = -q$, where $q \equiv |Q|$, the Hamiltonian operator for our system can be written as:
\begin{equation} \label{ogHamil}
    \mathcal{H} = \frac{\left(\mathbf{P} + q\textbf{A}\right)^{2}}{2\mu},
\end{equation}
where $\textbf{P} = -i \hbar \nabla$ is the momentum operator, and $\textbf{A}$ is the vector potential satisfying $\nabla \times \textbf{A} = \textbf{B}$. Here, the strategy is to recast our Hamiltonian into the form of a equivalent to the quantum harmonic oscillator. Cyclotron motion is dominated by motion in the plane perpendicular to the magnetic field ($|\textbf{k}_\perp| >> k_\parallel$, where $\parallel$ denotes the direction along the magnetic field). Therefore we will neglect $\pi_{z}$ in the following analysis. Let us define a pair of creation/annihilation operators, $a^\dag$ and $a$ respectively, in terms of the kinetic momentum operator $\mathbf{\pi} = \textbf{P} + q\textbf{A}$:
\begin{equation} \label{aDagA}
\begin{cases}
    a^\dag &= \frac{1}{\sqrt{2}}\frac{\delta}{\hbar} \left( \pi_{x} + i\pi_{y}\right) ,\\
    a &= \frac{1}{\sqrt{2}} \frac{\delta}{\hbar} \left( \pi_{x}- i\pi_{y}\right), \\
\end{cases}
\end{equation}
where $\delta \equiv \sqrt{\frac{\hbar}{qB_{0}}}$ is the magnetic length. Given that $\pi_{x}$ and $\pi_{y}$ satisfy $[\pi_{x},\pi_{y}] = -i\frac{\hbar^2}{\delta^2}$ it is easy to show that $[a,a^\dag] = 1$. It is also straight forward to prove that this pair of creation/annihilation operators possess the usual properties that:
\begin{align}
a^\dag a \psi_{n} &= n\psi_{n}  \label{nProp}, \\ 
a^\dag \psi_{n} &= \sqrt{n+1} ~\psi_{n+1} \label{aDagProp},\\
a \psi_{n} &= \sqrt{n} ~\psi_{n-1} \label{aProp}, 
\end{align}
where $n$ is the Landau level index. Defining the cyclotron frequency $\omega_{c} \equiv \frac{qB_{0}}{\mu}$, we can equivalently write Eq.\,\eqref{ogHamil} as:
\begin{equation} \label{hamiltonian}
    \mathcal{H} = \hbar \omega_{c} \left( a^\dag a + \frac{1}{2} \right). 
\end{equation}
In this work, we adopt the symmetric form of the vector potential for a uniform magnetic field $\textbf{A} = \frac{B_{0}}{2} \langle -y,x,0 \rangle $. Defining $w = x + iy$ and $\Bar{w} = x-iy$, we have:
\begin{equation} \label{aDagA_w}
\begin{cases}
    a^\dag = -i \sqrt{2} \delta \left[ \partial_{\Bar{w}} -  w / 4 \delta^{2}\right] ,\\
    a  = -i \sqrt{2} \delta \left[ \partial_{w} +  \Bar{w} / 4 \delta^{2} \right],\\
\end{cases}
\end{equation}
where we have used the fact that $\partial_{w} = \frac{\partial_{x} - i \partial_{y}}{2}$ and $\partial_{\Bar{w}} = \frac{\partial_{x} + i \partial_{y}}{2}$. Asserting that $a \psi(w,\bar{w}) = 0$ will furnish our ground state (i.e., $n = 0$) wave function by solving
\begin{equation}
-i \sqrt{2} \delta \left[ \partial_{w} +  \Bar{w} / 4 \delta^{2} \right] \psi(w,\bar{w}) = 0,
\end{equation}
which has solutions of the form $\psi = f(w)e^{-w\bar{w} / 4\delta^{2}}$, where $f(w)$ is an arbitrary holomorphic function of $w$. Physically, $f(w)$ captures the infinitely degenerate nature of the ground state wave function in this system. As such, we make the choice $f(w) = w^{m}$ for $m \in \mathbb{Z}$ (including 0). Asserting that $\psi$ be properly normalized allows us to derive our normalization factor, $N_{m}$, and express our set of degenerate ground states as:
\begin{align} 
    \psi_{n=0,m} &= N_{m} w^{m} e^{-w\bar{w} / 4\delta^{2}}, \\
    N_{m} & \equiv \left[ \pi m! \left(2 \delta^{2} \right)^{m+1}\right]^{-1/2}.
\end{align}
Using Eq.\,\eqref{aDagProp} we can produce the full spectrum of excited states as:
\begin{equation} \label{psi_nm_adag}
    \psi_{n,m} = \frac{\left( a^\dag \right)^{n}}{\sqrt{n!}} \psi_{0,m}.
\end{equation}
Combining Eqs.\,\eqref{nProp} and \eqref{hamiltonian}, we have that $\mathcal{H}\psi_{n,m} = \hbar \omega_{c} (n + 1/2) \psi_{n,m}$; in this sense we can understand that the Landau level index represents which energy level $n$ a given eigenstate occupies with degeneracy $m$. Combining Eqs.\,\eqref{hamiltonian} and \eqref{aDagA_w} we can express the Hamiltonian operator in terms of $(w,\bar{w})$ as:

\begin{equation}
    \mathcal{H} / \hbar \omega_{c} =  -2 \delta^{2} \partial_{w} \partial_{\bar{w}} - \frac{1}{2} \left( \bar{w}\partial_{\bar{w}}  - w \partial_{w} \right) + \frac{1}{8\delta^{2}}w\bar{w} ,
\end{equation}
where it is important to recognize the second term contains the canonical angular momentum operator ${L}_{z} = -i \hbar \partial_{\theta} = \hbar \left( w \partial_{w} - \bar{w} \partial_{\bar{w}} \right)$. In this sense each energy eigenstate of $\mathcal{H}$ can be understood to also be an eigenstate of $\textbf{L}_{z}$ and can therefore be recast in the form $\psi_{n,m} ~ \propto ~e^{-i(n-m)\theta}$. From here it is straight forward to prove that the $\psi_{n,m}$'s constitute an orthogonal basis for which $\langle \psi_{n,m} | \psi_{n',m'} \rangle = 0 ~ \forall (n,m) \neq (n',m')$.

To fully describe a given eigenstate $\psi$ it is evident that we require two quantum numbers, $n$ and $m$. Therefore, to ensure our desired coherent state fully captures the physics of the prescribed Hamiltonian we require a second pair of creation/annihilation operators. We turn our attention to a pair of creation/annihilation operators $\left(b^\dag,b \right)$ defined as:

\begin{equation}
    \begin{cases}
        b^\dag = \frac{1}{\sqrt{2}\delta} \left( x_g - iy_g \right),\\
        b = \frac{1}{\sqrt{2}\delta} \left( x_g + iy_g \right),\\
    \end{cases}
\end{equation}
where $x_g \equiv x - \pi_{y} / \mu \omega_{c}$ and $y_g \equiv y + \pi_{x} / \mu \omega_{c}$ are the guiding center position operators for an negatively charged particle. In terms of $\left(w,\bar{w}\right)$ these operators take the form:

\begin{equation}
    \begin{cases}
        b^\dag = \sqrt{2}\delta \left(-\partial_{{w}} + \bar{w} / 4 \delta^{2} \right),\\
        b = \sqrt{2}\delta \left( \partial_{\bar{w}} + w / 4 \delta^{2} \right).\\
\end{cases}
\end{equation}
As with $(a^\dag,a)$, we have that $[b,b^\dag] = 1$ and:
\begin{align}
b^\dag b \psi_{n,m} &= m\psi_{n,m}  , \\ 
b^\dag \psi_{n,m} &= \sqrt{m+1} ~\psi_{n,m+1} \label{bDagProp},\\
b \psi_{n,m} &= \sqrt{m} ~\psi_{n,m-1} , 
\end{align}
which allows us to express $\psi_{n,m}$ as:
\begin{equation}
    \psi_{n,m} = \frac{\left( b^\dag \right)^{m}}{\sqrt{m!}} \psi_{n,0}.
\end{equation}

The motion of a charged particle in a uniform magnetic field is fully prescribed by (1) the particle's angular momentum ($\ell_{z}$)  and (2) the particle's guiding center position $\left(x_{g},y_{g} \right)$. Whereas the information contained in the pair of quantum operators $(a^\dag,a)$ corresponds to classical information about a particle's angular momentum, the information contained in the pair of quantum operators $(b^\dag,b)$ corresponds to classical information about the guiding center of the particle's gyro-orbit.

\subsection{Coherent State Construction} \label{csSection}
Constructing a coherent state from the eigenstates derived in the previous subsection can be done in a manner similar to the case of the quantum harmonic oscillator. That is, a coherent state $\Psi$ can be formed as:

\begin{equation}
    \Psi = D(\alpha,\beta) \psi_{0,0},
\end{equation}
where $D\left( \alpha,\beta\right)$ is the displacement operator that is dependent on two parameters $\alpha$ and $\beta$.
Applying the displacement operator can be done efficiently by first proving that $[a^\dag,b^\dag] = [a^\dag,b] = [a,b^\dag] = [a,b] = [b^\dag,H] = [b,H] = 0$, which allows us to simply write:

\begin{equation} \label{generalCoherent}
    \Psi = e^{-\frac{|\alpha|^2 +  |\beta|^2}{2}} e^{\alpha a^\dag + \beta b^\dag} \psi_{0,0},
\end{equation}

For $\alpha = -i w_{0} / \sqrt{2}\delta$ and $\beta = \lambda_{0} / \sqrt{2}\delta$, with complex variables $w_{0}$ and $\lambda_{0}$, Eq.\,\eqref{generalCoherent} expands out to:
\begin{equation}
    \begin{split} \label{coherent_scratch}
        \Psi = & ~ e^{-\frac{|w_{0}|^2 + |\lambda_{0}|^2}{4\delta^2}} e^{-w_{0}\left[\partial_{\Bar{w}} - {w}/4\delta^{2} \right]} \left[e^{-\lambda_{0}\left[\partial_{{w}} - 
    \bar{w}/4\delta^{2} \right]} \right. \\
        & \left. \left(N_{0}e^{-w\bar{w} / 4\delta^2}\right)\right].
    \end{split}
\end{equation}

We can make use of the fact that $e^{c\frac{\partial}{\partial x}}f(x) = f(x+c)$ to simplify our operator application, and yield:

\begin{equation} \label{coherent_w}
        \Psi =  N_{0}e^{ -\frac{1}{4 \delta^2} \left[ |w|^2 + 2 \left(w_{0}\lambda_{0} - w_{0} {w} - \lambda_{0} \bar{w} \right) + |w_{0}|^2 + |\lambda_{0}|^2 \right]}.
\end{equation}
However, by rearranging $\text{Arg}\left[\Psi\right]$, one can recast Eq.\,\eqref{coherent_scratch} into a more physically illuminating form given by


\begin{equation} \label{coherent_xy}
    \begin{split}
        \Psi(x,y) = & N_{0} ~\text{exp} \left[-\frac{1}{4\delta^2} |x- x_{0}|^{2} - \frac{1}{4\delta^2}|y - y_{0}|^{2} \right. \\
        & \left. + ik_x x +ik_y y - i\varphi_{0} \right],
    \end{split}
\end{equation}

where,

\begin{equation} \label{constantDefs}
    \begin{split}
    x_{0} \equiv ~& \text{Re}[w_{0} + \lambda_{0}] ,\\
    y_{0} \equiv ~& \text{Im}[ \lambda_{0} - w_{0} ] ,\\
    k_{x} \equiv ~& \frac{1}{2 \delta^{2}} \text{Im}[\lambda_{0}+w_{0}] ,\\
    k_{y} \equiv ~& \frac{1}{2 \delta^{2}} \text{Re}[w_{0} - \lambda_{0} ] ,\\
    \varphi_{0} \equiv ~&
    \frac{1}{2 \delta^{2}}\text{Im}[\lambda_{0}w_{0}].\\
    \end{split}
\end{equation}
Such a coherent state has a probability distribution function given by

\begin{equation} \label{psiSqr}
    |\Psi(x,y)|^{2} = N_{0}^{2} e^{-\frac{1}{2\delta^2} |x- x_{0}|^{2} -\frac{1}{2\delta^2} |y - y_{0}|^{2}}.
\end{equation}
In this form, it is easy to understand that the probability distribution given by Eq.\,\eqref{psiSqr} describes a Gaussian wave packet of width $\delta$ (magnetic length), centered about the point $(x,y) = (x_0,y_0)$. 

\begin{figure}[h!]
    \centering
    \includegraphics[width=\linewidth]{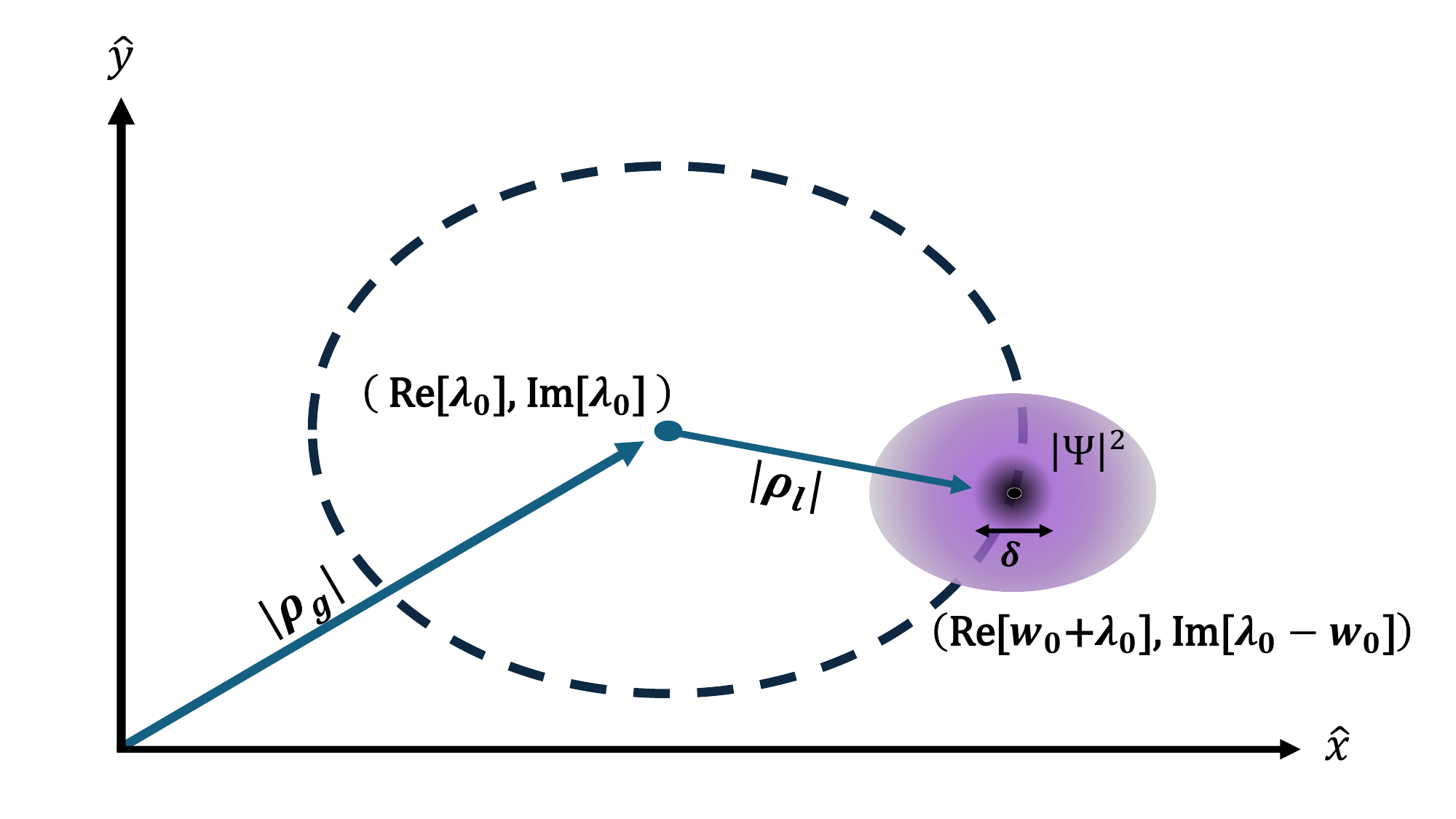}
    \caption{A qualitative depiction of the geometry of the system described by Eqs.\,\eqref{psiSqr} and \eqref{qmClassCorss}}
    \label{fig:coherentStateFigure}
\end{figure}

To interpret the physical meaning of the parameters $(w_{0},\lambda_{0})$ parameters it is straight forward to prove 
\begin{equation} \label{qmClassCorss}
    \begin{split} 
        \langle \Psi|\textbf{r}| \Psi\rangle =& (x_{0},y_{0})  , \\ 
        \langle \Psi|(x_{g},y_{g})| \Psi \rangle = &\left(\text{Re}\left[\lambda_{0}\right],\text{Im}\left[\lambda_{0}\right] \right),  \\
        \left(x_{\ell},y_{\ell}\right) \equiv& \langle \Psi|(x,y)| \Psi \rangle - \langle \Psi|(x_{g},y_{g})| \Psi \rangle \\
         = &  \left(\text{Re}\left[w_{0} \right],-\text{Im}\left[w_{0}\right] \right), \\
        \left(k_{x},k_{y}\right) =& \frac{1}{2 \delta^2} \left(y_g - y_\ell, x_\ell - x_g \right) , \\
    \end{split}
\end{equation}
where $\left(x_{\ell},y_{\ell}\right)$ correspond to the position of the charged particle on the cyclotron orbit. Therefore, the guiding center of the cyclotron orbit, $(x_{g},y_{g})$, is given by $(x_{g},y_{g}) = \left(\text{Re}\left[\lambda_{0}\right],\text{Im}\left[\lambda_{0}\right] \right)$ and the Larmor radius of the orbit, $\rho_{\ell}$ is given by $ \rho_{\ell} = |w_{0}|$. In Fig.\,\ref{fig:coherentStateFigure} we provide a sample depiction of the geometry of our system including a sample profile of $|\psi|^2$, the guiding center position, cyclotron orbit, and the Larmor radius.

We can compute the expected canonical angular momentum $\ell_{z} \equiv \langle \Psi |L_{z}| \Psi \rangle$,

\begin{align}
    \ell_{z} = & \hbar ~ \frac{|\rho_\ell|^2 - |\rho_{g}|^2}{2\delta^2} \label{lz_q},\\
    \ell_{z} = & \frac{1}{2}qB_{0} \left( \rho_{\ell}^2 - \rho_{g}^2 \right), \label{lz_c}
\end{align}
where $|\rho_{g}|$ is the guiding center radius. 
With these relations in hand it is straight forward to show that the global quantum energy partition of this system is given by 
\begin{align} 
\langle \Psi |\mathcal{H}| \Psi \rangle = & \frac{\hbar \omega_c}{2}+ \frac{1}{2} \mu \omega_{c}^2\rho_{\ell}^2 \label{expectH_split}\\
 = & \frac{1}{2} \mu \omega_{c}^2\rho_{\ell}^2 \left[1 + \frac{\delta^2}{\rho_{\ell}^2}\right] \label{expectH_class},
\end{align}
where the first term represents the vibrational kinetic energy stored in individual eigenmodes of our coherent state, and the second term represents the translational kinetic energy of the wave packet's gyro-orbit. Eq.\,\eqref{expectH_class} is simply a recasting of Eq.\,\eqref{expectH_split} that demonstrates that the total energy of the quantum subsystem is simply the classical kinetic energy of a particle with $\mu$ and velocity $\omega_c \rho_{\ell}$ modified by an order $\mathcal{O} \left( \frac{\delta^2}{\rho_{\ell}^2} \right)$ correction. In the classical limit of $\hbar \rightarrow 0$, we have that $\delta \rightarrow 0$ and we recover the expected classical energy. This expression also highlights the importance of quantum effects in high-field physics where the magnetic length and Larmor radius may be comparable. 



To obtain a time-dependent coherent state in the presence of static electromagnetic fields, it is useful to expand Eq.\,\eqref{generalCoherent} in Fock space as:
\begin{equation} \label{stationaryCoherent}
    \Psi = e^{-\frac{|\alpha|^2 + |\beta|^2}{2}} \sum_{n=0}^{\infty} \sum_{m=0}^{\infty} \frac{\left(\alpha a^\dag\right)^{n}}{\sqrt{n!}}\frac{\left(\beta b^\dag\right)^{m}}{\sqrt{m!}} \psi_{n,m}.
\end{equation}
Utilizing the fact that each eigenstate of $\mathcal{H}$ evolves as $\psi_{n,m}(t) = \psi_{n,m}(0) ~\text{exp}[-i\omega_{c}(n+1/2)]$, we can rewrite Eq.\,\eqref{stationaryCoherent} to derive the time-dependent coherent state, $\Psi(t)$ as:

\begin{equation} \label{}
    \begin{split}
        \Psi(t) =& e^{-\frac{|\alpha|^2 + |\beta|^2}{2}} \sum_{n=0}^{\infty} \sum_{m=0}^{\infty} \frac{\left(\alpha a^\dag\right)^{n}}{\sqrt{n!}}\frac{\left(\beta b^\dag\right)^{m}}{\sqrt{m!}} \psi_{n,m} e^{-i\omega_{c}(n+\frac{1}{2})t} ,\\
        \Psi(t) =& e^{-\frac{|\alpha|^2 + |\beta|^2 +i\omega_{c}t}{2}} \sum_{n=0}^{\infty} \sum_{m=0}^{\infty} \frac{\left(\alpha(t) a^\dag\right)^{n}}{\sqrt{n!}}\frac{\left(\beta b^\dag\right)^{m}}{\sqrt{m!}} \psi_{n,m} , \\
        \Psi(t) =& e^{-i\omega_{c}t/2} \Psi|_{\alpha = \alpha(t)},
    \end{split}
\end{equation}
where we use $\Psi|_{\alpha = \alpha(t)}$ to denote the time-independent coherent state, $\Psi$, being evaluated at $\alpha = \alpha e^{-i\omega_{c}t} \equiv \alpha(t)$. From the definition of $\alpha$, we can further define $w_{0}(t) \equiv w_{0}e^{-i\omega_{c}t} $ such that $\alpha(t) = -i w_{0}(t) / \sqrt{2}\delta$. This modifies Eq.\,\eqref{qmClassCorss} accordingly as

\begin{equation}
    \begin{split}
        (x_g,y_g)=& \left(\text{Re}\left[\lambda_{0}\right],\text{Im}\left[\lambda_{0}\right] \right)  ,\\
        x_{\ell}(t) =& ~\text{Re}[w_{0}]\text{cos}(\omega_{c}t) +\text{Im}[w_{0}]\text{sin}(\omega_{c}t),  \\
        y_{\ell}(t) =& ~ \text{Re}[w_{0}]\text{sin}(\omega_{c}t) - \text{Im}[w_{0}]\text{cos}(\omega_{c}t), \\
        (x_{0}(t),y_{0}(t)) = & ~ (x_{g} + x_{\ell}(t),~y_{g} + y_{\ell}(t)). 
    \end{split}
\end{equation}

As expected, only the orbital position of our charged particle relative to the guiding center evolves in time while the guiding center position itself remains fixed.

\bibliography{refs}

\end{document}